\documentclass[10pt,twocolumn,twoside]{IEEEtran}

\usepackage{graphicx}
\usepackage[cmex10]{amsmath}
\usepackage{amsfonts}
\usepackage{bm}
\usepackage{array}

\DeclareMathOperator{\re}{Re}
\DeclareMathOperator{\im}{Im}

\newcommand{\STF}{T_{\text{\rm STF}}}
\newcommand{\NTF}{T_{\text{\rm NTF}}}
\newcommand{\I}{{\mathrm{I}}}
\newcommand{\C}{{\mathbb{C}}}
\newcommand{\tr}{^\top}
\newcommand{\cl}{\text{cl}}
\newcommand{\OSR}{N_{\text{OSR}}}
\newcommand{\IB}{{\boldsymbol{I}}_{\text{B}}}
\newcommand{\jj}{\mathrm{j}}
\newcommand{\ee}{\mathrm{e}}
\newcommand{\dd}{\mathrm{d}}
\newcommand{\Va}{{\mathcal N}_{\text{average}}}
\newcommand{\Vw}{{\mathcal N}_{\text{worst}}}

\newcommand{\Ew}{{\mathcal E}_{\text{worst}}}
\newcommand{\SNR}{\mathrm{SNR}_{\text{pp}}}
\newcommand{\Ilow}{{\boldsymbol{I}}_{\text{low}}}
\newcommand{\Imid}{{\boldsymbol{I}}_{\text{mid}}}
\newcommand{\Il}{{\boldsymbol{I}}_l}
\newcommand{\Jlow}{J_{\text{low}}}
\newcommand{\Jmid}{J_{\text{mid}}}
\newcommand{\Jmb}{J_{\text{mb}}}

\newtheorem{defn}{Definition}
\newtheorem{lemma}{Lemma}
\newtheorem{prop}{Proposition}
\newtheorem{rem}{Remark}
\newtheorem{ass}{Assumption}
\newtheorem{theorem}{Theorem}
\newtheorem{cor}{Corollary}
\newtheorem{problem}{Problem}

\begin{document}
%
\title{Frequency Domain Min-Max Optimization of Noise-Shaping Delta-Sigma Modulators}
%

\author{
\thanks{
	Copyright (c) 2012 IEEE. 
	Personal use of this material is permitted. 
	However, permission to use this material for any other purposes must be obtained 
	from the IEEE by sending a request to pubs-permissions@ieee.org.
}
	Masaaki~Nagahara,~\IEEEmembership{Member,~IEEE,}
        Yutaka~Yamamoto,~\IEEEmembership{Fellow,~IEEE}%
\thanks{M. Nagahara and Y. Yamamoto are with Graduate School of
Informatics,
Kyoto University, Kyoto 606-8501, Japan.}
\thanks{e-mail: nagahara@ieee.org (\mbox{Nagahara}), yy@i.kyoto-u.ac.jp (\mbox{Yamamoto})}%
}

\markboth{IEEE TRANSACTIONS ON SIGNAL PROCESSING,~Vol.~60, No.~6, JUNE~2012}%
{NAGAHARA AND YAMAMOTO: Frequency Domain Min-Max Optimization of Noise-Shaping Delta-Sigma Modulators}

\maketitle

\begin{abstract}
This paper proposes a min-max design of
noise-shaping delta-sigma ($\Delta\Sigma$) modulators.
We first characterize the all stabilizing loop-filters
for a linearized modulator model.
By this characterization,
we formulate the design problem of lowpass, bandpass, and
multi-band modulators
as minimization of
the maximum magnitude of the noise transfer function (NTF)
in fixed frequency band(s).
We show that this optimization minimizes the worst-case reconstruction error,
and hence improves the SNR (signal-to-noise ratio) of the modulator.
The optimization is reduced to an optimization 
with a linear matrix inequality (LMI)
via the generalized KYP (Kalman-\hspace{0pt}Yakubovich-\hspace{0pt}Popov) lemma.
The obtained NTF is an FIR (finite-impulse-response) filter,
which is favorable in view of implementation.
We also derive a stability condition for the nonlinear model of $\Delta\Sigma$ modulators
with general quantizers including uniform ones.
This condition is described as an $H^\infty$ norm condition,
which is reduced to an LMI via the KYP lemma.
Design examples show advantages of our design.
\end{abstract}
\begin{IEEEkeywords}
Delta-sigma modulators, min-max optimization, noise-shaping, quantization.
\end{IEEEkeywords}


\IEEEpeerreviewmaketitle

\section{Introduction}
\label{sec:introduction}

\IEEEPARstart{D}{elta sigma}
($\Delta\Sigma$, see Table~\ref{tbl:abbreviations} on the next page
for the list of acronyms) modulators are widely used
in oversampling AD (Analog-to-Digital) and DA (Digital-to-Analog) converters,
by which we can achieve high performance with coarse quantizers~\cite{NorSchTem,SchTem}.
They have applications in digital signal processing systems,
such as digital audio~\cite{JanRee03,Zol} and
digital communications~\cite{VleRabWoo01,RusDonIsm06,GusEriFag10}.
More recently, the notion of $\Delta\Sigma$ modulators is
extended to several research areas related to signal processing.
In~\cite{DauDeVor03,BenPowYil06,LamPowYil10},
the $\Delta\Sigma$ scheme is introduced
for quantizing coefficients in finite but redundant frame
expansion of signals,
and is proved to outperform the standard PCM (pulse code modulation) scheme.
Based on this study, $\Delta\Sigma$ scheme is 
also applied to compressed sensing~\cite{BouBar07,GunLamPowSaaYlm10}.
In~\cite{AzuSug07,AzuSug08}, dynamic quantizers as $\Delta\Sigma$ modulators
are proposed for controlling linear time-invariant systems with
discrete-valued control inputs.
The $\Delta\Sigma$ scheme is also applied to obtain an approximate solution
of large discrete quadratic programming problems~\cite{CalBizRovSet10}.
For independent source separation~\cite{FazGorCha10}
and manifold learning~\cite{GorCha10},
machine learning is combined with $\Sigma\Delta$ modulation,
called the \emph{$\Sigma\Delta$ learner}.

In designing $\Delta\Sigma$ modulators,
noise shaping is a fundamental issue~\cite{SchTem}.
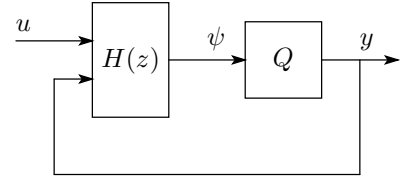
\begin{figure}[t]
\centering
\unitlength 0.1in
\begin{picture}( 20.0000,  9.2000)(  4.0000,-11.0000)
%
\special{pn 8}%
\special{pa 400 400}%
\special{pa 800 400}%
\special{fp}%
\special{sh 1}%
\special{pa 800 400}%
\special{pa 734 380}%
\special{pa 748 400}%
\special{pa 734 420}%
\special{pa 800 400}%
\special{fp}%
%
\special{pn 8}%
\special{pa 800 200}%
\special{pa 1200 200}%
\special{pa 1200 800}%
\special{pa 800 800}%
\special{pa 800 200}%
\special{fp}%
%
\special{pn 8}%
\special{pa 1200 500}%
\special{pa 1600 500}%
\special{fp}%
\special{sh 1}%
\special{pa 1600 500}%
\special{pa 1534 480}%
\special{pa 1548 500}%
\special{pa 1534 520}%
\special{pa 1600 500}%
\special{fp}%
%
\special{pn 8}%
\special{pa 1600 300}%
\special{pa 2000 300}%
\special{pa 2000 700}%
\special{pa 1600 700}%
\special{pa 1600 300}%
\special{fp}%
%
\special{pn 8}%
\special{pa 2000 500}%
\special{pa 2400 500}%
\special{fp}%
\special{sh 1}%
\special{pa 2400 500}%
\special{pa 2334 480}%
\special{pa 2348 500}%
\special{pa 2334 520}%
\special{pa 2400 500}%
\special{fp}%
%
\special{pn 8}%
\special{pa 2200 500}%
\special{pa 2200 1100}%
\special{fp}%
\special{pa 2200 1100}%
\special{pa 600 1100}%
\special{fp}%
\special{pa 600 1100}%
\special{pa 600 600}%
\special{fp}%
%
\special{pn 8}%
\special{pa 600 600}%
\special{pa 800 600}%
\special{fp}%
\special{sh 1}%
\special{pa 800 600}%
\special{pa 734 580}%
\special{pa 748 600}%
\special{pa 734 620}%
\special{pa 800 600}%
\special{fp}%
\put(10.0000,-5.0000){\makebox(0,0){$H(z)$}}%
\put(18.0000,-5.0000){\makebox(0,0){$Q$}}%
\put(22.0000,-4.5000){\makebox(0,0)[lb]{$y$}}%
\put(14.0000,-4.5000){\makebox(0,0)[lb]{$\psi$}}%
\put(4.0000,-3.5000){\makebox(0,0)[lb]{$u$}}%
\end{picture}%
\caption{$\Delta\Sigma$ modulator with loop-filter $H=[H_1,H_2]$ and quantizer $Q$.}
\label{fig:delta-sigma}
\end{figure}%
To describe the issue of noise shaping,
let us consider a general $\Delta\Sigma$ modulator 
shown in Fig.~\ref{fig:delta-sigma}.
In this figure, $Q$ is a quantizer and $H=[H_1, H_2]$ 
is a linear filter with 2 inputs and 1 output.
The filter $H_1$ shapes the signal transfer function (STF)
from the input $u$ to the output $y$
to have a unity magnitude in the frequency band of interest.
On the other hand, the filter $H_2$ eliminates the in-band quantization noise
by shaping the noise transfer function (NTF).
Then, if the input signal $u$ is sufficiently oversampled,
the frequency components in $u$ are concentrated in the band of interest,
and hence one can effectively extract the original signal $u$
from the quantized signal $y$ by applying a lowpass filter
to $y$ with a suitable cutoff frequency.
In fact, it is theoretically shown that the reconstruction error decreases
rapidly as the oversampling ratio increases~\cite{DauDeVor03, KraWar12}.

A usual solution to noise shaping is to insert accumulators (or integrators)
in the feedback loop
to attenuate the magnitude of the NTF in low frequency. 
To improve upon the performance, accumulators are cascaded
in various ways such as 
the MASH (multi-stage noise-shaping) modulators~\cite{HayInaUchiIwa86,CanHuy86}.
This methodology is analogous to a PID (Proportional-Integral-Derivative)
control~\cite{DatHoBha}, 
in which the performance of the designed system depends on 
the experience of the designer.
That is, the conventional design is of ad hoc nature.

To obtain a systematic design method,
one can adopt a more general type of transfer functions than accumulators 
for $H(z)$ in Fig.~\ref{fig:delta-sigma}.
From this point of view, 
the NTF zero optimization~\cite{Sch93,SchTem} was proposed
to shape the NTF \emph{optimally} in the frequency band of
interest, say $[0,\Omega]$.  
This optimization is done by changing the zeros of the NTF
so as to minimize the normalized noise power
given by the integral of the squared magnitude of the NTF
over $[0,\Omega]$.
While this method gives a systematic way to design $\Delta\Sigma$ modulators,
it can yield a peak in the magnitude of the NTF at a certain frequency, 
since such a peak cannot be captured by an 
integrated or averaged objective function.
A recent paper~\cite{Ho+06} has investigated this problem
and proposed to use semi-infinite programming for constraining the maximum value
of a function over the frequency band.
This method, however, does not necessarily 
optimize
the overall performance but only minimizes
the denominator of a loop-filter.
That is, the method~\cite{Ho+06} does not necessarily
reduce peaks in the NTF magnitude.
Also, the computational cost for the optimization
is very high due to its infinite dimensionality.
Alternatively, the present authors has
proposed to adopt $H^\infty$ optimization for attenuating the
NTF magnitude itself with a frequency-domain weighting function~\cite{NagWadYam06}.
This method gives a good performance \emph{if} a suitable
weighting function was chosen.
For general notion of $H^\infty$ optimization
in signal processing,
see~\cite{Nag11,YamNagKha12}.
The well-known Remez exchange method (aka Parks-McClellan method)~\cite{ParMcCle72}
is related to the $H^\infty$ optimization.
The method gives a near-optimal filter that minimizes the
maximum error between a given desired filter and the filter to be
designed.
Strictly speaking, this is not $H^\infty$-optimal since
the response is ignored on the transition frequency band.

In contrast to these methods, we propose%
\footnote{This method was first proposed in our conference articles~\cite{NagYam09-1,NagYam09-2}.
The present paper organizes these works with new results on
SNR performance (Section~\ref{subsec:worstcase}),
bandpass modulator design (Section~\ref{subsec:gkyp_bp}),
and stability theorems (Section~\ref{sec:stability}).
Simulation results in Section~\ref{sec:examples}
are also new.} 
a novel design based on \emph{min-max} optimization,
which can be reduced to finite dimensional convex optimization.
That is, we directly \emph{minimize the maximum magnitude} of the NTF over
the frequency band of interest.
In other words, 
we design $\Delta\Sigma$ modulators in order to uniformly attenuate the magnitude
over the prespecified band.
This uniform minimization
improves the \emph{worst-case}
SNR (signal-to-noise ratio)
to be defined in Section~\ref{subsec:worstcase}, 
of the modulator in the band of interest.
Conversely, a peak of the NTF magnitude as above can deteriorate the 
worst-case SNR 
and also the dynamic range of the modulator.
We propose in this paper a more effective method
that does not require a selection of a weighting function. 

To this end, we first characterize all stabilizing loop-filters 
for a linearized modulator.  Then, by using this parametrization, 
we formulate the design problem as an optimization 
via a linear matrix inequality (LMI)
for lowpass and bandpass modulators 
using the generalized Kalman-\hspace{0pt}Yakubovich-\hspace{0pt}Popov 
(KYP) lemma~\cite{IwaHar05,Nag11}. 
Furthermore, we can assign arbitrarily zeros of the NTF 
on the unit circle in the complex plane
by adding a linear matrix equality (LME)
constraint to the LMI.
These techniques 
are mostly adopted from \emph{control theory}.
Recently, control theory is effectively applied to $\Delta\Sigma$
modulator design
with finite horizon predictive control~\cite{QueGoo05,OstQueJen11},
sliding mode control~\cite{Yu06}, and
robust control~\cite{YanGan08,McKerGanYanHen09},
to name a few. In particular,
the idea of applying the generalized KYP lemma to $\Delta\Sigma$ modulator 
design is proposed in~\cite{OsqRooMeg07}, 
in which they assume a one-bit quantizer for $Q$ and optimize the
average power of the reconstruction error in low frequency
for lowpass modulators.  
In contrast, our approach minimizes 
the \emph{worst} reconstruction error,
which can improve the overall SNR as mentioned above.

Stability analysis of $\Delta\Sigma$ modulators is another fundamental issue.
For first-order~\cite{TepConFee05} and second-order~\cite{GunTha04,Ho+11}
modulators, stability is well-studied in terms of invariant set.
On the other hand,
we derive a stability condition taking account of nonlinearity in $\Delta\Sigma$
modulators of arbitrary order
with general quantizers including uniform ones.
This condition is derived in terms of a state-space representation,
and is described by the $\ell^1$ norm of a linear system.
This can be transformed into an $H^\infty$-norm condition of the NTF
as a sufficient condition.
This $H^\infty$-norm constraint can be equivalently expressed
as an LMI via the KYP lemma~\cite{BoyGhaFerBal,YamAndNagKoy03,Nag11}.
In summary, the proposed method can be described
by LMI's and LME's,
which can be solved effectively by numerical
optimization softwares 
such as YALMIP~\cite{Lof04} and SeDuMi~\cite{Stu01} with MATLAB.

The organization of this paper is as follows:
Section~\ref{sec:design} gives characterization of
all loop-filters that stabilize a linearized feedback modulator.
Section~\ref{sec:gkyp} is the main section of this paper,
in which we motivate the min-max design in view of SNR improvement,
and then we formulate the design 
as a min-max optimization, which is reduced to LMI's and LME's.
Section~\ref{sec:stability} discusses stability 
of the $\Delta\Sigma$ modulator model without linearization.
Section~\ref{sec:cascade} introduces a cascade structure
for high-order modulators.
Section~\ref{sec:examples} gives design examples
to show advantages of our method.
Section~\ref{sec:conclusion} concludes our study.

\subsection*{Notation and Convention}

Throughout this paper, we use the following notations.
Abbreviations in this paper are summed up in Table~\ref{tbl:abbreviations}.
\begin{description}
\item[${\mathcal{S}}$, ${\mathcal{S}}'$]
$\mathcal{S}$ is the set of all stable, causal, and rational transfer functions
with real coefficients, and
${\mathcal{S}}':=\{R\in{\mathcal{S}}: R \ \text{is strictly causal}\}$.
\item[$\ell^1$] the Banach space of all real-valued absolutely summable sequences.
For $\{v(k)\}_{k\geq 0}\in\ell^1$, the $\ell^1$ norm is defined by
$\|v\|_1 := \sum_{k\geq 0} |v(k)|$.
\item[$\ell^\infty$] the Banach space of all real-valued bounded sequences.
For $\{v(k)\}_{k\geq 0}\in\ell^\infty$, the $\ell^\infty$ norm is defined by
$\|v\|_\infty := \sup_{k\geq 0} |v(k)|$.
\item[$v\ast w$] convolution of two sequences 
$\{v(k)\}_{k\geq 0}$ and $\{w(k)\}_{k\geq 0}$, 
that is,
\[
 (v \ast w)(m) := \sum_{k\geq 0} v(m-k)w(k), \quad m=0,1,2,\dots.
\]
For this computation, we set $v(m-k)=0$ if $m<k$.
\end{description}
\begin{table}[t]
\caption{Abbreviations}
\label{tbl:abbreviations}
\centering
\begin{tabular}{|c|c|}\hline
abbrev. & full name \\\hline
$\Delta\Sigma$ & Delta Sigma\\
NTF &  Noise Transfer Function\\
STF &  Signal Transfer Function\\
OSR &  Over-Sampling Ratio\\
SNR &  Signal-to-Noise Ratio\\
KYP &  Kalman Yakubovich Popov\\
LMI &  Linear Matrix Inequality\\
LME &  Linear Matrix Equality\\\hline
\end{tabular}
\end{table}

\section{Characterization of Loop-Filters}
\label{sec:design}

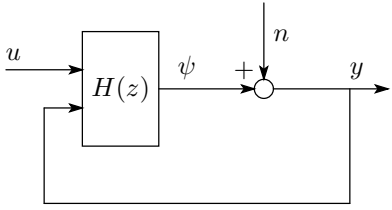
\begin{figure}[tbp]
\centering
\unitlength 0.1in
\begin{picture}( 20.0000, 10.5000)(  4.0000,-11.0000)
%
\special{pn 8}%
\special{pa 400 400}%
\special{pa 800 400}%
\special{fp}%
\special{sh 1}%
\special{pa 800 400}%
\special{pa 734 380}%
\special{pa 748 400}%
\special{pa 734 420}%
\special{pa 800 400}%
\special{fp}%
%
\special{pn 8}%
\special{pa 800 200}%
\special{pa 1200 200}%
\special{pa 1200 800}%
\special{pa 800 800}%
\special{pa 800 200}%
\special{fp}%
%
\special{pn 8}%
\special{pa 2200 500}%
\special{pa 2200 1100}%
\special{fp}%
\special{pa 2200 1100}%
\special{pa 600 1100}%
\special{fp}%
\special{pa 600 1100}%
\special{pa 600 600}%
\special{fp}%
%
\special{pn 8}%
\special{pa 600 600}%
\special{pa 800 600}%
\special{fp}%
\special{sh 1}%
\special{pa 800 600}%
\special{pa 734 580}%
\special{pa 748 600}%
\special{pa 734 620}%
\special{pa 800 600}%
\special{fp}%
\put(10.0000,-5.0000){\makebox(0,0){$H(z)$}}%
\put(22.0000,-4.5000){\makebox(0,0)[lb]{$y$}}%
\put(13.0000,-4.5000){\makebox(0,0)[lb]{$\psi$}}%
\put(4.0000,-3.5000){\makebox(0,0)[lb]{$u$}}%
%
\special{pn 8}%
\special{pa 1200 500}%
\special{pa 1700 500}%
\special{fp}%
\special{sh 1}%
\special{pa 1700 500}%
\special{pa 1634 480}%
\special{pa 1648 500}%
\special{pa 1634 520}%
\special{pa 1700 500}%
\special{fp}%
%
\special{pn 8}%
\special{ar 1750 500 50 50  0.0000000 6.2831853}%
%
\special{pn 8}%
\special{pa 1800 500}%
\special{pa 2400 500}%
\special{fp}%
\special{sh 1}%
\special{pa 2400 500}%
\special{pa 2334 480}%
\special{pa 2348 500}%
\special{pa 2334 520}%
\special{pa 2400 500}%
\special{fp}%
%
\special{pn 8}%
\special{pa 1750 50}%
\special{pa 1750 450}%
\special{fp}%
\special{sh 1}%
\special{pa 1750 450}%
\special{pa 1770 384}%
\special{pa 1750 398}%
\special{pa 1730 384}%
\special{pa 1750 450}%
\special{fp}%
\put(18.0000,-2.5000){\makebox(0,0)[lb]{$n$}}%
\put(17.0000,-4.5000){\makebox(0,0)[rb]{$+$}}%
\end{picture}%
\caption{Linearized model for $\Delta\Sigma$ modulator with loop-filter $H=[H_1,H_2]$.}
\label{fig:linear-modulator}
\end{figure}

In this section, we characterize all $H(z)$'s
that stabilize the linearized model shown in Fig.~\ref{fig:linear-modulator}.
This characterization is a basis for the proposed min-max design
formulated in Section~\ref{sec:gkyp}.
For a stability condition taking account of the nonlinear effect of the quantizer,
see the discussion in Section~\ref{sec:stability}.

We first define causality, stability, well-posedness and internal stability
of linear systems.
\begin{defn}[Causality and Stability]
A rational transfer function $P(z)$ is said to be \emph{(strictly) causal}
if the order of the numerator of $P(z)$ is 
(strictly) less than that of the denominator,
and said to be \emph{stable} if the poles of $P(z)$ are
all in the open unit disk ${\mathbb D}=\{z\in{\mathbb C}:|z\rvert <1\}$.
\end{defn}
\begin{defn}[Well-posedness]
The feedback system in Fig.~\ref{fig:linear-modulator} is \emph{well-posed} if
there is at least one clock of delay in $H_2(z)$,
that is, the transfer function $H_2(z)$ is strictly causal.
\end{defn}
\begin{defn}[Internal stability]
The feedback system Fig.~\ref{fig:linear-modulator} is \emph{internally stable} if
the four transfer functions from $[u, n]\tr$ to $[\psi, y]\tr$ are all stable.
\end{defn}

We here characterize the filter $H(z)$ that makes
the linearized feedback system well-posed and internally stable.
All stabilizing filters are characterized
as follows:
\begin{prop}
\label{prop:R}
The linearized feedback system in Fig.~\ref{fig:linear-modulator} is well-posed and 
internally stable if and only if
\begin{equation}
\begin{split}
H_1(z) &= \frac{P(z)}{1+R(z)},
\quad H_2(z) = \frac{R(z)}{1+R(z)},\\
P(z) &\in {\mathcal{S}}, 
\quad R(z) \in {\mathcal{S}}',
\end{split}
\label{eq:param}
\end{equation}
where ${\mathcal{S}}$ denotes the set of all stable, causal, 
and rational transfer functions with real coefficients,
and
${\mathcal{S}}':=\{R(z)\in{\mathcal{S}}: R(z) \ \text{is strictly causal}\}$.
\end{prop}
\begin{IEEEproof}
See Appendix~\ref{ap:lemma1}.
\end{IEEEproof}

By using the parameters $P(z)\in{\mathcal{S}}$ and $R(z)\in{\mathcal{S}}'$,
we obtain
the STF and NTF respectively as
$\STF(z)=P(z)$ and $\NTF(z)=1+R(z)$.
From this, it follows that the input/output equation of the system 
in Fig.~\ref{fig:linear-modulator}
is given by
\begin{equation}
y = \STF~u + \NTF~n = Pu + (1+R)n.
\label{eq:R}
\end{equation}

\begin{figure}[tbp]
\centering
\unitlength 0.1in
\begin{picture}( 27.0000, 14.2000)(  0.0000,-16.0000)
%
\special{pn 8}%
\special{pa 0 400}%
\special{pa 200 400}%
\special{fp}%
\special{sh 1}%
\special{pa 200 400}%
\special{pa 134 380}%
\special{pa 148 400}%
\special{pa 134 420}%
\special{pa 200 400}%
\special{fp}%
%
\special{pn 8}%
\special{pa 200 200}%
\special{pa 600 200}%
\special{pa 600 600}%
\special{pa 200 600}%
\special{pa 200 200}%
\special{fp}%
%
\special{pn 8}%
\special{pa 600 400}%
\special{pa 800 400}%
\special{fp}%
\special{sh 1}%
\special{pa 800 400}%
\special{pa 734 380}%
\special{pa 748 400}%
\special{pa 734 420}%
\special{pa 800 400}%
\special{fp}%
%
\special{pn 8}%
\special{ar 850 400 50 50  0.0000000 6.2831853}%
%
\special{pn 8}%
\special{pa 900 400}%
\special{pa 1500 400}%
\special{fp}%
\special{sh 1}%
\special{pa 1500 400}%
\special{pa 1434 380}%
\special{pa 1448 400}%
\special{pa 1434 420}%
\special{pa 1500 400}%
\special{fp}%
%
\special{pn 8}%
\special{pa 1500 200}%
\special{pa 1900 200}%
\special{pa 1900 600}%
\special{pa 1500 600}%
\special{pa 1500 200}%
\special{fp}%
%
\special{pn 8}%
\special{pa 1900 400}%
\special{pa 2700 400}%
\special{fp}%
\special{sh 1}%
\special{pa 2700 400}%
\special{pa 2634 380}%
\special{pa 2648 400}%
\special{pa 2634 420}%
\special{pa 2700 400}%
\special{fp}%
%
\special{pn 8}%
\special{pa 2300 400}%
\special{pa 2300 1000}%
\special{fp}%
%
\special{pn 8}%
\special{pa 2300 1000}%
\special{pa 1900 1000}%
\special{fp}%
\special{sh 1}%
\special{pa 1900 1000}%
\special{pa 1968 1020}%
\special{pa 1954 1000}%
\special{pa 1968 980}%
\special{pa 1900 1000}%
\special{fp}%
%
\special{pn 8}%
\special{pa 1200 400}%
\special{pa 1200 1000}%
\special{fp}%
%
\special{pn 8}%
\special{pa 1200 1000}%
\special{pa 1800 1000}%
\special{fp}%
\special{sh 1}%
\special{pa 1800 1000}%
\special{pa 1734 980}%
\special{pa 1748 1000}%
\special{pa 1734 1020}%
\special{pa 1800 1000}%
\special{fp}%
%
\special{pn 8}%
\special{ar 1850 1000 50 50  0.0000000 6.2831853}%
%
\special{pn 8}%
\special{pa 1850 1050}%
\special{pa 1850 1400}%
\special{fp}%
%
\special{pn 8}%
\special{pa 1850 1400}%
\special{pa 1650 1400}%
\special{fp}%
\special{sh 1}%
\special{pa 1650 1400}%
\special{pa 1718 1420}%
\special{pa 1704 1400}%
\special{pa 1718 1380}%
\special{pa 1650 1400}%
\special{fp}%
%
\special{pn 8}%
\special{pa 1650 1200}%
\special{pa 1250 1200}%
\special{pa 1250 1600}%
\special{pa 1650 1600}%
\special{pa 1650 1200}%
\special{fp}%
%
\special{pn 8}%
\special{pa 1250 1400}%
\special{pa 850 1400}%
\special{fp}%
%
\special{pn 8}%
\special{pa 850 1400}%
\special{pa 850 450}%
\special{fp}%
\special{sh 1}%
\special{pa 850 450}%
\special{pa 830 518}%
\special{pa 850 504}%
\special{pa 870 518}%
\special{pa 850 450}%
\special{fp}%
\put(4.0000,-4.0000){\makebox(0,0){$P(z)$}}%
\put(17.0000,-4.0000){\makebox(0,0){$Q$}}%
\put(14.5000,-14.0000){\makebox(0,0){$R(z)$}}%
\put(23.5000,-3.5000){\makebox(0,0)[lb]{$y=Q\psi$}}%
\put(12.5000,-3.5000){\makebox(0,0)[lb]{$\psi$}}%
\put(0.5000,-3.5000){\makebox(0,0)[lb]{$u$}}%
\put(8.0000,-3.5000){\makebox(0,0)[rb]{$+$}}%
\put(8.0000,-4.5000){\makebox(0,0)[rt]{$+$}}%
\put(18.0000,-9.5000){\makebox(0,0)[rb]{$-$}}%
\put(19.0000,-9.5000){\makebox(0,0)[lb]{$+$}}%
\put(19.0000,-12.5000){\makebox(0,0)[lb]{$n$}}%
\end{picture}%
\caption{Error-feedback structure of $\Delta\Sigma$ modulator
with design parameters  $P(z)\in\mathcal{S}$ and $R(z)\in\mathcal{S}'$.}
\label{fig:mod2}
\end{figure}
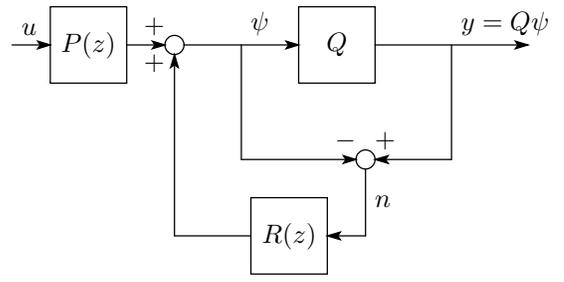

By equation \eqref{eq:R}, the $\Delta\Sigma$ modulator can be realized by means of 
the design parameters $P(z)\in\mathcal{S}$ and  $R(z)\in\mathcal{S}'$
as shown in Fig.~\ref{fig:mod2}.
This structure, called \emph{error-feedback structure}~\cite{SchTem}
or \emph{noise-shaping coder}~\cite{NorSchTem},
is often applied in the digital loops required in 
$\Delta\Sigma$ DA converters~\cite{SchTem}.
By this block diagram,
we can interpret the filter $P(z)$ as a pre-filter 
to shape the frequency response of the input signal,
and $R(z)$ as a feedback gain for the quantization noise $Q\psi-\psi$.

\section{Optimal Loop-Filter Design via Linear Matrix Inequalities and Equalities}
\label{sec:gkyp}

In this section,
we propose a min-max design of the loop-filter $H(z)$
by using the parametrization in Proposition~\ref{prop:R}.
First, we introduce the worst-case analysis of reconstruction errors in
$\Delta\Sigma$ modulators to motivate the min-max design to be proposed.
We then present design procedures for lowpass and bandpass modulators.

\subsection{Worst-case analysis of reconstruction errors}
\label{subsec:worstcase}

In oversampling lowpass $\Delta\Sigma$ converters
with \emph{oversampling ratio} $\OSR$ (see Table~\ref{tbl:abbreviations})~\cite{SchTem},
the authors attempt to attenuate the magnitude of the NTF in 
the frequency band  $\IB=[0,\Omega]\subset[0,\pi]$
 where $\Omega=\pi/\OSR$.
In a bandpass converter, the band will be 
$\IB=[\omega_0-\Omega,\omega_0+\Omega]$
where $\omega_0\in(0,\pi)$ is the center frequency.
We here consider a general interval $\IB\subset[0,\pi]$ in which
the magnitude of the NTF is designed to be small.
In a conventional design~\cite{Sch93,SchTem}, the attenuation level 
of the magnitude is measured by the \emph{average}
or the \emph{root mean square}
\begin{equation}
 \Va(\NTF,\IB):=\sqrt{\frac{1}{\lvert \IB\rvert }\int_{\IB} 
  \lvert \NTF(\ee^{\jj\omega})\rvert ^2 \dd\omega}.
 \label{eq:Va}
\end{equation}
On the other hand, we consider the \emph{worst-case} measure
\begin{equation}
 \Vw(\NTF,\IB):=\max_{\omega\in\IB}\left\lvert \NTF(\ee^{\jj\omega})\right\rvert .
 \label{eq:Vw}
\end{equation}
It is easy to see that $\Vw$ gives an upper bound of $\Va$, that is,
\[
 \Va(\NTF,\IB) \leq \Vw(\NTF,\IB).
\]
Hence, minimization of $\Vw(\NTF,\IB)$ leads to small $\Va(\NTF,\IB)$,
but not conversely.  One can give an NTF with the same average $\Va$ 
but much larger maximum magnitude $\Vw$.
That is, a small $\Va$ does not necessarily yield a small $\Vw$.

Another advantage of minimizing $\Vw$ is the worst-case optimization 
of the reconstruction error $y-u$
(see Fig.~\ref{fig:mod2}).
Define the \emph{worst-case reconstruction error} $\Ew$ by
\[
 \Ew := \max_{\omega\in\IB} \left\lvert \hat{y}(\ee^{\jj\omega})-\hat{u}(\ee^{\jj\omega})\right\rvert ,
\]
where $\hat{y}$ and $\hat{u}$ are, respectively, the discrete-time Fourier transforms 
of $y$ and $u$ in Fig.~\ref{fig:mod2}.
Then this quantity can be described by the maximum magnitude 
$\Vw(\NTF,\IB)$
of $\NTF(z)$ over $\IB$.
In fact, we have the following proposition:
\begin{prop}
\label{prop:SNR}
Assume that the magnitude $\lvert \hat{n}(\jj\omega)\rvert $ 
of the quantization noise $n=Q\psi-\psi$
is bounded on $\IB$,
that is, there exists $C_0>0$ such that
$\max_{\omega\in\IB} \lvert \hat{n}(\ee^{\jj\omega})\rvert  = C_0$.
Assume also that
\begin{equation}
 \lvert \STF(\ee^{\jj\omega})\rvert =1,\quad \forall \omega\in\IB.
 \label{eq:STF_gain}
\end{equation}
Then the worst-case reconstruction error is given by
\begin{equation}
 \Ew = C_0 \cdot \Vw(\NTF,\IB).
 \label{eq:Eworst}
\end{equation}
\end{prop}
\begin{IEEEproof}
By the relation
\[
 \begin{split}
 \hat{y}(\ee^{\jj\omega}) &= \STF(\ee^{\jj\omega}) \hat{u}(\ee^{\jj\omega}) 
  + \NTF(\ee^{\jj\omega}) \hat{n}(\ee^{\jj\omega})\\
  &= \hat{u}(\ee^{\jj\omega}) + \NTF(\ee^{\jj\omega}) \hat{n}(\ee^{\jj\omega}),\quad
 \forall \omega\in\IB,
 \end{split}
\]
we have
\[
 \lvert \hat{y}(\ee^{\jj\omega})-\hat{u}(\ee^{\jj\omega})\rvert  
  = \lvert \NTF(\ee^{\jj\omega})\hat{n}(\ee^{\jj\omega})\rvert,\quad  \forall \omega\in\IB.
\]
By taking the maximum over the interval $\IB$, we obtain \eqref{eq:Eworst}.
\end{IEEEproof}
Note that the assumption \eqref{eq:STF_gain} holds if we choose
the pre-filter $P(z)$ that has a unity magnitude response over $\IB$.
In particular, if we take $P(z)=1$ then we have $\STF(z)=1$.
By Proposition 2,
optimization of $\Vw$ improves the worst-case reconstruction error $\Ew$.
Minimizing $\Vw$ also improves
the peak-to-peak SNR (signal-to-noise ratio) of the modulator
defined by
\begin{equation}
 \SNR(u) 
 := 
 \frac{
   \max_{\omega\in\IB}\lvert \hat{u}(\ee^{\jj\omega})\rvert^2 }
 {
   \max_{\omega\in\IB}\lvert \hat{y}(\ee^{\jj\omega})-\hat{u}(\ee^{\jj\omega})\rvert^2 }.
 \label{eq:SNRworst}
\end{equation}
\begin{figure}[t]
\centering
\unitlength 0.1in
\begin{picture}( 30.9000, 20.7300)(  3.1000,-22.5500)
%
\special{pn 8}%
\special{pa 400 2200}%
\special{pa 400 200}%
\special{fp}%
\special{sh 1}%
\special{pa 400 200}%
\special{pa 380 268}%
\special{pa 400 254}%
\special{pa 420 268}%
\special{pa 400 200}%
\special{fp}%
%
\special{pn 8}%
\special{pa 400 2200}%
\special{pa 3400 2200}%
\special{fp}%
\special{sh 1}%
\special{pa 3400 2200}%
\special{pa 3334 2180}%
\special{pa 3348 2200}%
\special{pa 3334 2220}%
\special{pa 3400 2200}%
\special{fp}%
%
\special{pn 20}%
\special{pa 2502 2020}%
\special{pa 2472 2002}%
\special{pa 2442 1988}%
\special{pa 2412 1980}%
\special{pa 2382 1982}%
\special{pa 2354 1994}%
\special{pa 2296 2026}%
\special{pa 2266 2040}%
\special{pa 2234 2046}%
\special{pa 2202 2050}%
\special{pa 2170 2050}%
\special{pa 2136 2050}%
\special{pa 2102 2050}%
\special{pa 2070 2044}%
\special{pa 2040 2036}%
\special{pa 2010 2024}%
\special{pa 1984 2004}%
\special{pa 1938 1958}%
\special{pa 1912 1940}%
\special{pa 1884 1936}%
\special{pa 1852 1940}%
\special{pa 1818 1954}%
\special{pa 1788 1972}%
\special{pa 1762 1996}%
\special{pa 1738 2018}%
\special{pa 1714 2036}%
\special{pa 1688 2042}%
\special{pa 1660 2038}%
\special{pa 1630 2026}%
\special{pa 1600 2010}%
\special{pa 1568 1992}%
\special{pa 1536 1978}%
\special{pa 1502 1968}%
\special{pa 1470 1964}%
\special{pa 1438 1966}%
\special{pa 1406 1972}%
\special{pa 1376 1984}%
\special{pa 1344 1998}%
\special{pa 1314 2014}%
\special{pa 1284 2026}%
\special{pa 1252 2034}%
\special{pa 1222 2030}%
\special{pa 1162 2004}%
\special{pa 1132 2002}%
\special{pa 1104 2014}%
\special{pa 1074 2030}%
\special{pa 1046 2044}%
\special{pa 1016 2056}%
\special{pa 984 2062}%
\special{pa 952 2064}%
\special{pa 916 2060}%
\special{pa 876 2050}%
\special{pa 838 2038}%
\special{pa 800 2030}%
\special{pa 770 2028}%
\special{pa 748 2040}%
\special{pa 738 2066}%
\special{pa 736 2104}%
\special{pa 734 2134}%
\special{pa 722 2144}%
\special{pa 702 2130}%
\special{pa 672 2104}%
\special{pa 640 2070}%
\special{pa 606 2038}%
\special{pa 574 2018}%
\special{pa 548 2016}%
\special{pa 530 2038}%
\special{pa 510 2062}%
\special{pa 482 2068}%
\special{pa 448 2064}%
\special{pa 410 2054}%
\special{pa 402 2050}%
\special{sp}%
%
\special{pn 20}%
\special{pa 2500 2030}%
\special{pa 2504 1886}%
\special{pa 2504 1840}%
\special{pa 2506 1792}%
\special{pa 2506 1744}%
\special{pa 2508 1696}%
\special{pa 2508 1650}%
\special{pa 2512 1466}%
\special{pa 2514 1420}%
\special{pa 2516 1332}%
\special{pa 2518 1246}%
\special{pa 2520 1162}%
\special{pa 2522 1082}%
\special{pa 2524 1004}%
\special{pa 2526 930}%
\special{pa 2528 860}%
\special{pa 2528 828}%
\special{pa 2530 764}%
\special{pa 2532 734}%
\special{pa 2532 704}%
\special{pa 2534 676}%
\special{pa 2534 650}%
\special{pa 2536 626}%
\special{pa 2536 602}%
\special{pa 2538 580}%
\special{pa 2538 558}%
\special{pa 2540 538}%
\special{pa 2540 520}%
\special{pa 2542 504}%
\special{pa 2542 488}%
\special{pa 2544 476}%
\special{pa 2544 464}%
\special{pa 2546 454}%
\special{pa 2546 446}%
\special{pa 2548 440}%
\special{pa 2548 434}%
\special{pa 2550 432}%
\special{pa 2550 430}%
\special{pa 2552 432}%
\special{pa 2552 434}%
\special{pa 2554 438}%
\special{pa 2554 446}%
\special{pa 2556 454}%
\special{pa 2556 464}%
\special{pa 2558 474}%
\special{pa 2558 488}%
\special{pa 2560 504}%
\special{pa 2560 520}%
\special{pa 2562 538}%
\special{pa 2562 558}%
\special{pa 2564 578}%
\special{pa 2564 600}%
\special{pa 2566 624}%
\special{pa 2566 650}%
\special{pa 2568 676}%
\special{pa 2568 704}%
\special{pa 2570 732}%
\special{pa 2570 762}%
\special{pa 2572 794}%
\special{pa 2572 826}%
\special{pa 2574 894}%
\special{pa 2576 966}%
\special{pa 2578 1002}%
\special{pa 2580 1080}%
\special{pa 2582 1160}%
\special{pa 2584 1286}%
\special{pa 2588 1418}%
\special{pa 2588 1464}%
\special{pa 2590 1510}%
\special{pa 2590 1554}%
\special{pa 2592 1602}%
\special{pa 2592 1648}%
\special{pa 2594 1742}%
\special{pa 2596 1790}%
\special{pa 2596 1836}%
\special{pa 2600 2028}%
\special{pa 2600 2030}%
\special{sp}%
%
\special{pn 20}%
\special{pa 2594 2024}%
\special{pa 2632 2000}%
\special{pa 2666 1982}%
\special{pa 2694 1974}%
\special{pa 2716 1986}%
\special{pa 2726 2016}%
\special{pa 2734 2050}%
\special{pa 2746 2076}%
\special{pa 2768 2080}%
\special{pa 2798 2066}%
\special{pa 2832 2042}%
\special{pa 2868 2018}%
\special{pa 2904 1998}%
\special{pa 2934 1992}%
\special{pa 2958 2008}%
\special{pa 2974 2042}%
\special{pa 2992 2054}%
\special{pa 3014 2036}%
\special{pa 3040 2006}%
\special{pa 3066 1992}%
\special{pa 3092 2004}%
\special{pa 3114 2032}%
\special{pa 3134 2064}%
\special{sp}%
%
\special{pn 8}%
\special{pa 400 1926}%
\special{pa 3130 1926}%
\special{da 0.070}%
%
\special{pn 8}%
\special{pa 2550 432}%
\special{pa 400 432}%
\special{da 0.070}%
\put(4.8000,-3.1200){\makebox(0,0)[lb]{$\lvert\hat{y}(\ee^{\jj\omega})\rvert$ in dB}}%
%
\special{pn 8}%
\special{pa 1874 1086}%
\special{pa 1874 1926}%
\special{fp}%
\special{sh 1}%
\special{pa 1874 1926}%
\special{pa 1894 1860}%
\special{pa 1874 1874}%
\special{pa 1854 1860}%
\special{pa 1874 1926}%
\special{fp}%
%
\special{pn 8}%
\special{pa 1874 1086}%
\special{pa 1874 432}%
\special{fp}%
\special{sh 1}%
\special{pa 1874 432}%
\special{pa 1854 500}%
\special{pa 1874 486}%
\special{pa 1894 500}%
\special{pa 1874 432}%
\special{fp}%
\put(18.1000,-9.9000){\makebox(0,0)[rb]{$\SNR(u)\approx$}}%
%
\special{pn 8}%
\special{pa 3126 2054}%
\special{pa 3126 2200}%
\special{dt 0.045}%
\put(31.2600,-23.2000){\makebox(0,0){$\Omega$}}%
\put(4.0000,-23.2000){\makebox(0,0){$0$}}%
\put(32.9000,-21.6000){\makebox(0,0)[lb]{$\omega$}}%
\end{picture}%
\caption{Peak-to-peak SNR for a narrow-band signal.}
\label{fig:SNR}
\end{figure}
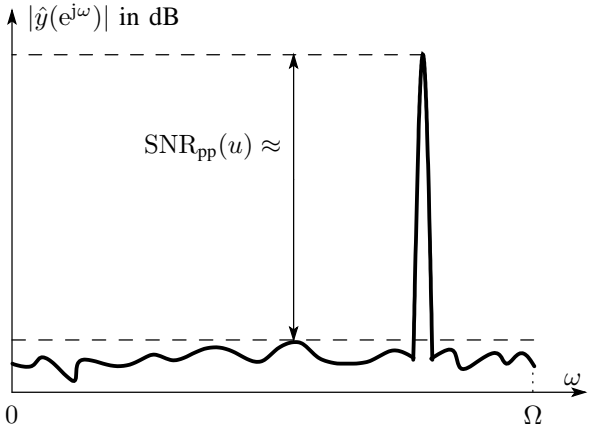%
Let us consider the following set of input signals:
\[
 {\mathcal U} := \left\{u: \max_{\omega\in\IB} \lvert \hat{u}(\ee^{\jj\omega})\rvert^2 = 1 \right\}.
\]
Suppose that the assumptions in Proposition \ref{prop:SNR}
hold.
Then, by Proposition \ref{prop:SNR},
we have
\[
 \mathrm{SNR}_{\text{worst}}:=\min_{u\in{\mathcal U}} \SNR(u) = \frac{1}{C_0\Vw(\NTF,\IB)}.
\]
It follows that smaller $\Vw$ leads to better \emph{worst-case} SNR.
Note that if condition \eqref{eq:STF_gain} holds and if the input signal is sufficiently
narrow-banded,
$\SNR$ can be estimated by the difference%
\footnote{
The difference is also known as the 
\emph{spurious-free dynamic range} (SFDR).} 
between the peak of $\hat{y}(\jj\omega)$ 
and the peak of noise, or the maximum noise level in $\lvert \hat{y}(\ee^{\jj\omega})\rvert$,
over the frequency range $\IB$
(see Fig.~\ref{fig:SNR}).

Conversely, if $\Ew$ is as large as 
$\max_{\omega\in\IB} \lvert \hat{u}(\ee^{\jj\omega})\rvert $,
then the $\SNR$ will be very poor,
and the dynamic range will also be very narrow.
As seen above, minimizing $\Va$ can yield a large NTF magnitude at a certain frequency,
and hence the performance may be degraded.  
See examples in Section~\ref{sec:examples} where we illustrate that 
minimizing $\Vw$ improves the $\SNR$ better than minimizing $\Va$.

In what follows, we set $P(z)=1$ for simplicity,
and show design methods of the loop-filter $R(z)$.
Since the STF and the NTF can be designed \emph{independently}
by relation \eqref{eq:R}, 
one can design $P(z)$ after obtaining the loop-filter $R(z)$
such that $|P(\ee^{\jj\omega})| = 1$ over $\IB$
and $|P(\ee^{\jj\omega})|\ll 1$ over $[0,\pi]\setminus\IB$
to achieve better reconstruction performance.

\subsection{Min-max design of lowpass modulators}
\label{subsec:gkyp}

\begin{figure}[tbp]
\centering
\unitlength 0.1in
\begin{picture}( 29.7000, 14.1500)(  0.3000,-18.8500)
%
\special{pn 8}%
\special{pa 600 1800}%
\special{pa 600 600}%
\special{fp}%
\special{sh 1}%
\special{pa 600 600}%
\special{pa 580 668}%
\special{pa 600 654}%
\special{pa 620 668}%
\special{pa 600 600}%
\special{fp}%
%
\special{pn 8}%
\special{pa 600 1800}%
\special{pa 3000 1800}%
\special{fp}%
\special{sh 1}%
\special{pa 3000 1800}%
\special{pa 2934 1780}%
\special{pa 2948 1800}%
\special{pa 2934 1820}%
\special{pa 3000 1800}%
\special{fp}%
%
\special{pn 20}%
\special{pa 600 1700}%
\special{pa 624 1662}%
\special{pa 646 1624}%
\special{pa 668 1590}%
\special{pa 692 1558}%
\special{pa 714 1532}%
\special{pa 736 1512}%
\special{pa 758 1500}%
\special{pa 782 1496}%
\special{pa 804 1502}%
\special{pa 826 1518}%
\special{pa 850 1544}%
\special{pa 872 1572}%
\special{pa 894 1604}%
\special{pa 918 1636}%
\special{pa 940 1664}%
\special{pa 962 1686}%
\special{pa 986 1698}%
\special{pa 1008 1700}%
\special{pa 1030 1688}%
\special{pa 1054 1668}%
\special{pa 1076 1640}%
\special{pa 1098 1608}%
\special{pa 1120 1576}%
\special{pa 1144 1546}%
\special{pa 1166 1522}%
\special{pa 1188 1504}%
\special{pa 1212 1500}%
\special{pa 1234 1506}%
\special{pa 1256 1522}%
\special{pa 1280 1544}%
\special{pa 1302 1574}%
\special{pa 1324 1604}%
\special{pa 1348 1636}%
\special{pa 1370 1666}%
\special{pa 1392 1692}%
\special{pa 1416 1712}%
\special{pa 1438 1726}%
\special{pa 1460 1734}%
\special{pa 1482 1736}%
\special{pa 1504 1732}%
\special{pa 1528 1722}%
\special{pa 1548 1708}%
\special{pa 1570 1688}%
\special{pa 1592 1666}%
\special{pa 1612 1638}%
\special{pa 1634 1608}%
\special{pa 1654 1574}%
\special{pa 1674 1536}%
\special{pa 1692 1496}%
\special{pa 1712 1454}%
\special{pa 1730 1410}%
\special{pa 1746 1364}%
\special{pa 1764 1316}%
\special{pa 1780 1268}%
\special{pa 1796 1218}%
\special{pa 1810 1170}%
\special{pa 1824 1120}%
\special{pa 1838 1072}%
\special{pa 1850 1024}%
\special{pa 1862 980}%
\special{pa 1876 938}%
\special{pa 1888 900}%
\special{pa 1902 868}%
\special{pa 1914 838}%
\special{pa 1930 816}%
\special{pa 1944 800}%
\special{pa 1960 790}%
\special{pa 1976 788}%
\special{pa 1994 796}%
\special{pa 2014 812}%
\special{pa 2034 836}%
\special{pa 2056 864}%
\special{pa 2078 896}%
\special{pa 2102 928}%
\special{pa 2124 956}%
\special{pa 2148 980}%
\special{pa 2172 996}%
\special{pa 2196 1000}%
\special{pa 2220 994}%
\special{pa 2242 976}%
\special{pa 2266 950}%
\special{pa 2288 918}%
\special{pa 2310 886}%
\special{pa 2334 854}%
\special{pa 2356 828}%
\special{pa 2378 808}%
\special{pa 2400 800}%
\special{pa 2422 806}%
\special{pa 2446 820}%
\special{pa 2468 844}%
\special{pa 2490 874}%
\special{pa 2512 906}%
\special{pa 2536 936}%
\special{pa 2558 966}%
\special{pa 2582 988}%
\special{pa 2604 1002}%
\special{pa 2626 1006}%
\special{pa 2650 998}%
\special{pa 2672 984}%
\special{pa 2694 962}%
\special{pa 2716 934}%
\special{pa 2740 902}%
\special{pa 2762 866}%
\special{pa 2784 828}%
\special{pa 2800 800}%
\special{sp}%
\put(6.5000,-6.0000){\makebox(0,0)[lb]{$|\NTF(\ee^{\jj\omega})|$}}%
\put(29.5000,-17.5000){\makebox(0,0)[lb]{$\omega$}}%
\put(17.0000,-19.5000){\makebox(0,0){$\Omega$}}%
\put(5.5000,-13.5000){\makebox(0,0)[rb]{$\gamma$}}%
\put(6.0000,-19.5000){\makebox(0,0){$0$}}%
\put(28.0000,-19.5000){\makebox(0,0){$\pi$}}%
%
\special{pn 8}%
\special{pa 2800 1800}%
\special{pa 2800 800}%
\special{dt 0.045}%
%
\special{pn 8}%
\special{pa 1750 1800}%
\special{pa 1750 1350}%
\special{dt 0.045}%
\special{pa 1750 1350}%
\special{pa 600 1350}%
\special{dt 0.045}%
\end{picture}%
\caption{Min-max optimization of the lowpass NTF 
  in the frequency domain: minimize the maximum magnitude $\gamma$
  in the band $\Ilow=[0,\Omega]$.}
\label{fig:prob}
\end{figure}
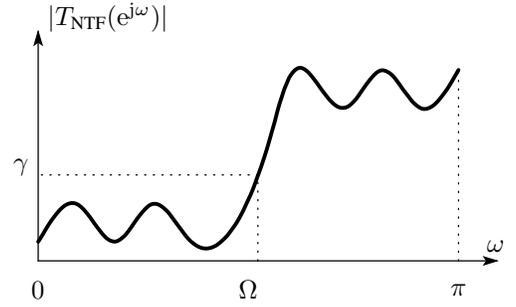

We now consider the design of lowpass modulators based on the discussion 
given in the previous section.  
Our objective here is
to find the loop-filter $R(z)$ that minimizes 
the magnitude of the frequency response of $\NTF(z)$
over $\Ilow:=[0,\Omega]$ as shown in Fig.~\ref{fig:prob}.
Our problem is formulated as follows:
\begin{problem}[Lowpass modulator]
Given $\Omega\ (0<\Omega<\pi)$,
find  $R(z)\in\mathcal{S}'$ that solves the following
min-max optimization: 
\[
\begin{split}
 \Jlow&:=\min_{R(z)\in\mathcal{S}'} \Vw(\NTF,\Ilow)\\
 &=\min_{R(z)\in\mathcal{S}'} \max_{\omega\in[0,\Omega]} |\NTF(\ee^{\jj\omega})|,
\end{split}
\]
or equivalently, 
\begin{gather}
 \text{minimize } \gamma \text{ subject to } R(z)\in\mathcal{S}'\text{ and}\nonumber\\
 \max_{\omega\in[0,\Omega]} |\NTF(\ee^{\jj\omega})|<\gamma.\label{eq:hinf}
\end{gather} 
\end{problem}

To solve this problem, we assume that $R(z)$ is 
a finite impulse response (FIR) filter, that is,
we set
\begin{equation}
R(z) = \sum_{k=0}^{N} \alpha_kz^{-k},\quad \alpha_0=0.
\label{eq:FIR}
\end{equation}
Note that the constraint $\alpha_0=0$ ensures $R(z)\in\mathcal{S}'$.
Note also that 
FIR filters are often preferred to IIR filters that may cause instability
attributed to quantization and recursion when they are implemented in
digital devices.
Therefore, the assumption to use FIR filter for $R(z)$ is not too restrictive.
We then introduce a state-space realization $\{A, B, C(\alpha)\}$, such that 
$R(z) = C(\alpha)(zI-A)^{-1}B$,
where $\alpha := [\alpha_0,\alpha_1,\dots,\alpha_N]$,
\begin{equation}
\begin{split}
A &:= \begin{bmatrix}0&1& &0\\
	 &\ddots&\ddots& \\
	 & &\ddots&1\\
	0& & &0
	\end{bmatrix},\quad
B := \begin{bmatrix}0\\\vdots\\0\\1\end{bmatrix},\\
C(\alpha) &:= [\alpha_N,\alpha_{N-1},\dots,\alpha_1].
\end{split}
\label{eq:R-ABC}
\end{equation}
Then inequality \eqref{eq:hinf} can be described as a linear matrix
inequality (LMI) by using the generalized KYP lemma~\cite{IwaHar05}: 
\begin{theorem}
\label{thm:gkyp}
Inequality \eqref{eq:hinf} holds if and only if
there exist symmetric matrices $Y>0$ and $X$ such that
\begin{equation}
\left[
\begin{array}{ccc}
M_{1}(X,Y) & M_{2}(X,Y) & C(\alpha)\tr\\
M_{2}(X,Y)\tr & M_{3}(X,\gamma^2) & 1\\
C(\alpha) & 1 & -1\\
\end{array}
\right]<0,
\label{eq:gkyp}
\end{equation}
where
\[
\begin{split}
 M_{1}(X,Y)&= A\tr XA + YA + A\tr Y - X - 2Y \cos\Omega,\\
 M_{2}(X,Y) &= A\tr XB + YB,\\
 M_{3}(X,\gamma^2) &= B\tr XB - \gamma^2.
\end{split}
\]
\end{theorem}
\begin{IEEEproof}
By the generalized KYP lemma~\cite[Theorem 2]{IwaHar05}
for the low frequency range $\Ilow=[0,\Omega]$
in the discrete-time setting, inequality \eqref{eq:hinf} is equivalent to
\[
 \left[
 \begin{array}{cc}
  M_1 & M_2\\ 
  M_2\tr & M_3
 \end{array}
 \right]
 +
 \bigl[
 \begin{array}{cc}
  C(\alpha)&1
 \end{array}
 \bigr]\tr
 \bigl[
 \begin{array}{cc}
  C(\alpha)&1
 \end{array}
 \bigr] < 0.
\]
Then applying the Schur complement~\cite[Sec.~2.1]{BoyGhaFerBal}
to this inequality
gives inequality \eqref{eq:gkyp}.
\end{IEEEproof}
By Theorem~\ref{thm:gkyp}, the optimal coefficients 
$\alpha_1, \dots \alpha_{N}$
of the filter $R(z)$ in \eqref{eq:FIR}
are obtained by minimizing $\gamma$ subject to \eqref{eq:gkyp}.
This LMI optimization is a convex optimization problem~\cite{BoyGhaFerBal,BoyVan},
and hence can be efficiently solved
by standard optimization softwares
e.g., MATLAB\@.
For optimization softwares and MATLAB codes, see Appendix~\ref{ap:matlab}.
\begin{rem}
The obtained NTF $\NTF(z)=1+R(z)$ is an FIR filter,
which is more preferred in view of implementation.  
On the other hand, a conventional optimal design~\cite{Sch93,SchTem}
yields an IIR (infinite-impulse-response) filter 
that has a problem of stability 
in digital implementation.
This is an advantage of the proposed design.
\end{rem}

\subsection{Min-max design of bandpass modulators}
\label{subsec:gkyp_bp}

Bandpass modulators are used in digital demodulation of 
frequency modulated analog signals, e.g., 
\cite{JanSchSne91}, \cite{JanMarSed97}.  

We can formulate the bandpass modulator design 
as a min-max optimization in the same light of lowpass modulators. 
Fig.~\ref{fig:prob_bp} illustrates noise shaping for bandpass modulators,
where $\omega_0 \in (0,\pi)$ is the center frequency and $2\Omega$ is
the bandwidth of interest.  
Our objective here is to minimize the magnitude of the NTF
over the frequency band
$\Imid:=[\omega_0-\Omega,\omega_0+\Omega]$.
Our design process is formulated as follows:
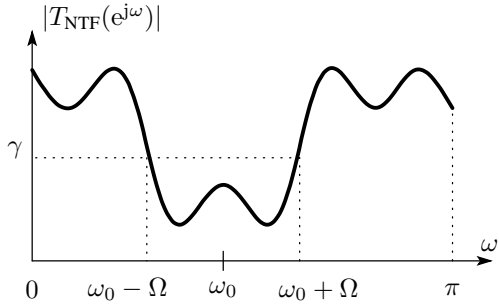
\begin{figure}[tbp]
\centering
\unitlength 0.1in
\begin{picture}( 29.7000, 14.1500)( -1.7000,-18.8500)
%
\special{pn 8}%
\special{pa 400 1800}%
\special{pa 400 600}%
\special{fp}%
\special{sh 1}%
\special{pa 400 600}%
\special{pa 380 668}%
\special{pa 400 654}%
\special{pa 420 668}%
\special{pa 400 600}%
\special{fp}%
%
\special{pn 8}%
\special{pa 400 1800}%
\special{pa 2800 1800}%
\special{fp}%
\special{sh 1}%
\special{pa 2800 1800}%
\special{pa 2734 1780}%
\special{pa 2748 1800}%
\special{pa 2734 1820}%
\special{pa 2800 1800}%
\special{fp}%
\put(4.5000,-6.0000){\makebox(0,0)[lb]{$|\NTF(\ee^{\jj\omega})|$}}%
\put(27.5000,-17.5000){\makebox(0,0)[lb]{$\omega$}}%
\put(19.0000,-19.5000){\makebox(0,0){$\omega_0+\Omega$}}%
\put(3.5000,-12.6000){\makebox(0,0)[rb]{$\gamma$}}%
\put(4.0000,-19.5000){\makebox(0,0){$0$}}%
\put(26.0000,-19.5000){\makebox(0,0){$\pi$}}%
%
\special{pn 8}%
\special{pa 2600 1800}%
\special{pa 2600 1000}%
\special{dt 0.045}%
%
\special{pn 20}%
\special{pa 400 800}%
\special{pa 422 838}%
\special{pa 444 874}%
\special{pa 468 906}%
\special{pa 490 938}%
\special{pa 512 962}%
\special{pa 534 984}%
\special{pa 558 996}%
\special{pa 580 1002}%
\special{pa 604 1000}%
\special{pa 628 988}%
\special{pa 652 968}%
\special{pa 676 942}%
\special{pa 698 914}%
\special{pa 722 884}%
\special{pa 744 856}%
\special{pa 766 830}%
\special{pa 786 810}%
\special{pa 806 798}%
\special{pa 826 792}%
\special{pa 844 796}%
\special{pa 860 806}%
\special{pa 876 822}%
\special{pa 890 844}%
\special{pa 906 872}%
\special{pa 918 906}%
\special{pa 932 942}%
\special{pa 944 982}%
\special{pa 958 1026}%
\special{pa 970 1072}%
\special{pa 982 1120}%
\special{pa 992 1168}%
\special{pa 1028 1316}%
\special{pa 1040 1362}%
\special{pa 1052 1406}%
\special{pa 1066 1450}%
\special{pa 1078 1488}%
\special{pa 1092 1522}%
\special{pa 1106 1552}%
\special{pa 1122 1576}%
\special{pa 1136 1596}%
\special{pa 1152 1606}%
\special{pa 1170 1612}%
\special{pa 1188 1608}%
\special{pa 1208 1594}%
\special{pa 1228 1574}%
\special{pa 1250 1548}%
\special{pa 1272 1518}%
\special{pa 1294 1486}%
\special{pa 1318 1458}%
\special{pa 1342 1432}%
\special{pa 1366 1412}%
\special{pa 1390 1402}%
\special{pa 1414 1402}%
\special{pa 1438 1414}%
\special{pa 1462 1434}%
\special{pa 1486 1460}%
\special{pa 1508 1488}%
\special{pa 1532 1520}%
\special{pa 1554 1550}%
\special{pa 1574 1576}%
\special{pa 1594 1596}%
\special{pa 1614 1608}%
\special{pa 1632 1612}%
\special{pa 1650 1606}%
\special{pa 1666 1594}%
\special{pa 1680 1576}%
\special{pa 1696 1552}%
\special{pa 1710 1522}%
\special{pa 1724 1486}%
\special{pa 1736 1448}%
\special{pa 1750 1406}%
\special{pa 1762 1360}%
\special{pa 1774 1314}%
\special{pa 1786 1264}%
\special{pa 1798 1214}%
\special{pa 1808 1164}%
\special{pa 1832 1066}%
\special{pa 1844 1020}%
\special{pa 1856 976}%
\special{pa 1870 936}%
\special{pa 1882 898}%
\special{pa 1896 864}%
\special{pa 1912 838}%
\special{pa 1926 814}%
\special{pa 1942 800}%
\special{pa 1958 790}%
\special{pa 1976 790}%
\special{pa 1996 798}%
\special{pa 2016 814}%
\special{pa 2036 838}%
\special{pa 2058 866}%
\special{pa 2080 896}%
\special{pa 2104 928}%
\special{pa 2128 956}%
\special{pa 2150 978}%
\special{pa 2174 994}%
\special{pa 2198 1000}%
\special{pa 2222 994}%
\special{pa 2244 978}%
\special{pa 2268 954}%
\special{pa 2290 926}%
\special{pa 2334 864}%
\special{pa 2358 836}%
\special{pa 2380 814}%
\special{pa 2402 800}%
\special{pa 2424 796}%
\special{pa 2446 802}%
\special{pa 2470 816}%
\special{pa 2492 838}%
\special{pa 2514 866}%
\special{pa 2538 898}%
\special{pa 2560 934}%
\special{pa 2582 970}%
\special{pa 2600 1000}%
\special{sp}%
\put(9.0000,-19.5000){\makebox(0,0){$\omega_0-\Omega$}}%
\put(14.0000,-19.5000){\makebox(0,0){$\omega_0$}}%
%
\special{pn 8}%
\special{pa 1400 1750}%
\special{pa 1400 1850}%
\special{fp}%
%
\special{pn 8}%
\special{pa 1800 1800}%
\special{pa 1800 1260}%
\special{dt 0.045}%
\special{pa 1800 1260}%
\special{pa 400 1260}%
\special{dt 0.045}%
%
\special{pn 8}%
\special{pa 1000 1260}%
\special{pa 1000 1800}%
\special{dt 0.045}%
\end{picture}%
\caption{Min-max optimization of the bandpass NTF in the frequency domain:
minimize the maximum magnitude $\gamma$
in the band $\Imid=[\omega_0-\Omega,\omega_0+\Omega]$.}
\label{fig:prob_bp}
\end{figure}%
\begin{problem}[Bandpass modulator]
Given $\omega_0 \in (0,\pi)$ and $\Omega>0$
such that 
$\Imid=[\omega_0-\Omega,\omega_0+\Omega] \subset [0,\pi]$,
find $R(z)\in{\mathcal{S}}'$ that solves the following
min-max optimization:
\[
\begin{split}
 \Jmid &:= \min_{R(z)\in\mathcal{S}'}\Vw(\NTF,\Imid)\\
 &=\min_{R(z)\in\mathcal{S}'}\max_{\omega\in[\omega_0-\Omega,\omega_0+\Omega]} 
   |\NTF(\ee^{\jj\omega})|,
\end{split}
\]
or equivalently,
\begin{gather}
 \text{minimize } \gamma 
 \text{ subject to } R(z)\in\mathcal{S}'\text{ and}\nonumber\\
 \max_{\omega\in[\omega_0-\Omega,\omega_0+\Omega]} |\NTF(\ee^{\jj\omega})|<\gamma.
 \label{eq:hinf_bp}
\end{gather} 
\end{problem}
As in the lowpass modulator design, 
we here constrain $R(z)$ to be an FIR filter defined in \eqref{eq:FIR}.
Let $\{A,B,C(\alpha)\}$ be state-space matrices as defined in the previous section.
Then the bandpass modulator problem is also reducible to an LMI optimization
via the generalized KYP lemma~\cite{IwaHar05}.
\begin{theorem}
\label{thm:gkyp_bp}
Inequality \eqref{eq:hinf_bp} holds if and only if 
there exist symmetric matrices $Y>0$ and $X$ such that
\begin{equation}
\left[
\begin{array}{ccc}
M_{4}(X,Y,\omega_0,\Omega) & M_{5}(X,Y,\omega_0) & C(\alpha)\tr\\
\overline{M}_{5}(X,Y,\omega_0)\tr & M_{6}(X,\gamma^2) & 1\\
C(\alpha) & 1 & -1\\
\end{array}
\right]<0,
\label{eq:gkyp_bp}
\end{equation}
where
\begin{equation}
\begin{split}
 M_{4}(X,Y,\omega_0,\Omega)&:= A\tr XA + YA\ee^{-\jj\omega_0}
   + A\tr Y\ee^{\jj\omega_0}\\&\qquad - X - 2Y \cos\Omega,\\
 M_{5}(X,Y,\omega_0) &:= A\tr XB + YB\ee^{-\jj\omega_0},\\
 \overline{M}_{5}(X,Y,\omega_0) &:= A\tr XB + YB\ee^{\jj\omega_0},\\
 M_{6}(X,\gamma^2) &:= B\tr XB - \gamma^2.
\end{split}
\label{eq:BPmat}
\end{equation}
\end{theorem}
\begin{IEEEproof}
By the generalized KYP lemma~\cite[Theorem 2]{IwaHar05}
for the mid frequency range $\Imid:=[\omega_0-\Omega,\omega_0+\Omega]$
in the discrete-time setting, 
inequality \eqref{eq:hinf_bp} is equivalent to
\[
 \left[
 \begin{array}{cc}
  M_4 & M_5\\ 
  \overline{M}_5\tr & M_6
 \end{array}
 \right]
 +
 \bigl[
 \begin{array}{cc}
  C(\alpha)&1
 \end{array}
 \bigr]\tr
 \bigl[
 \begin{array}{cc}
  C(\alpha)&1
 \end{array}
 \bigr] < 0.
\]
Then applying the Schur complement~\cite[Sec.~2.1]{BoyGhaFerBal}
to this inequality
gives inequality \eqref{eq:gkyp_bp}.
\end{IEEEproof}
\begin{rem}
LMI \eqref{eq:gkyp_bp} is complex-valued,
however, for some LMI solvers, a real-valued LMI is required.
An equivalent real-valued LMI for \eqref{eq:gkyp_bp} is given in
Appendix~\ref{ap:gkyp_bp}.
\end{rem}
\begin{rem} 
LMI \eqref{eq:gkyp_bp} with the center frequency $\omega_0=0$ 
is equivalent to LMI \eqref{eq:gkyp} for lowpass modulator.  
That is, Theorem~\ref{thm:gkyp} can be obtained as a special 
case of Theorem~\ref{thm:gkyp_bp}.  
\end{rem}

Theorem~\ref{thm:gkyp_bp} can be directly extended to
the following multi-band bandpass modulator design:
\begin{problem}[Multi-band bandpass modulator]
\label{prob:mb}
Given $\omega_l \in (0,\pi)$ and $\Omega_l>0$,
$l=1,2,\dots,L$
such that
\[
\Il=[\omega_l-\Omega_l,\omega_l+\Omega_l] \subset [0,\pi],~
l=1,2,\dots,L,
\]
find $R(z)\in{\mathcal{S}}'$ that solves the following
min-max optimization:
\[
\begin{split}
 \Jmb &:= \min_{R(z)\in\mathcal{S}'}\sum_{l=1}^L\Vw(\NTF,\Il)^2\\
 &=\min_{R(z)\in\mathcal{S}'}\sum_{l=1}^L \max_{\omega\in[\omega_l-\Omega_l,\omega_l+\Omega_l]} 
   |\NTF(\ee^{\jj\omega})|^2,
\end{split}
\]
or equivalently,
\begin{gather}
 \text{minimize } \gamma_1^2+\dots+\gamma_L^2 \text{ subject to }  R(z)\in\mathcal{S}' \text{ and}\nonumber\\
  \max_{\omega\in[\omega_l-\Omega_l,\omega_l+\Omega_l]} |\NTF(\ee^{\jj\omega})|<\gamma_l,~l=1,2,\dots,L.
  \label{eq:hinf_mb}
\end{gather}
\end{problem}
\begin{theorem}
\label{thm:gkyp_mb}
Inequalities \eqref{eq:hinf_mb} hold if and only if
there exist symmetric matrices $Y_l>0$ and $X_l$,
$l=1,2,\dots,L$ such that
\begin{align}
&\left[
\begin{array}{ccc}
M_{4}(X_l,Y_l,\omega_l,\Omega_l) & M_{5}(X_l,Y_l,\omega_l) & C(\alpha)\tr\\
\overline{M}_{5}(X_l,Y_l,\omega_l)\tr & M_{6}(X_l,\gamma_l^2) & 1\\
C(\alpha) & 1 & -1\\
\end{array}
\right]<0,\label{eq:gkyp_mb}\\
&l=1,2,\dots,L,\nonumber
\end{align}
where $M_4$, $M_5$, $\overline{M}_5$, and $M_6$ are
defined in \eqref{eq:BPmat}.
\end{theorem}
\begin{IEEEproof}
A direct consequence of Theorem~\ref{thm:gkyp_bp}.
\end{IEEEproof}

\subsection{NTF zeros}
\label{subsec:zeros}

To ensure perfect reconstruction of the DC input level,
and to reduce low-frequency tones,
$\NTF(z)$ should have zeros at $z=1$,
or the frequency $\omega=0$~\cite{SchTem}.
A similar requirement is for a bandpass $\Delta\Sigma$ modulator;
we set NTF zeros at a given frequency $\omega_0\in (0,\pi)$, 
or $z=\ee^{\pm \jj\omega_0}$.
The zeros of $\NTF(z)$ can be assigned by
linear equations (linear constraints) of $\alpha_1,\dots,\alpha_N$.
Define $\nu(z):=z^N + \sum_{k=1}^{N} \alpha_kz^{N-k}$.
Then, $\NTF(z)$ has $\mu$ zeros at $z=z_0$ if and only if
\[
\left.\frac{d^k \nu(z)}{dz^k}\right|_{z=z_0} = 0, \quad k=0,1,\dots,\mu-1,
\]
where $\frac{d^0\nu(z)}{dz^0}:=\nu(z)$.
The LMI with these linear constraints can also be effectively solved.

\section{Stability of Nonlinear Feedback Systems}
\label{sec:stability}

Although the linearized model in Fig.~\ref{fig:linear-modulator} is useful for analyzing
and designing noise-shaping $\Delta\Sigma$ modulators as above,
the stability of $\Delta\Sigma$ modulators should be analyzed 
with respect to their nonlinear behaviors induced by the quantizer $Q$. 
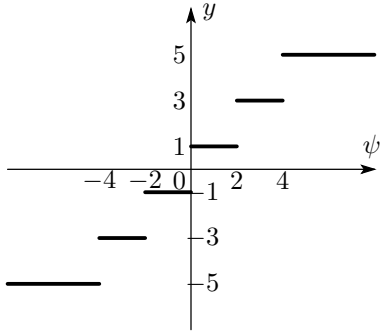
\begin{figure}[tbp]
\centering
\unitlength 0.1in
\begin{picture}( 19.2000, 17.9000)(  0.4000,-18.4000)
%
\special{pn 8}%
\special{pa 40 1000}%
\special{pa 1960 1000}%
\special{fp}%
\special{sh 1}%
\special{pa 1960 1000}%
\special{pa 1894 980}%
\special{pa 1908 1000}%
\special{pa 1894 1020}%
\special{pa 1960 1000}%
\special{fp}%
%
\special{pn 8}%
\special{pa 1000 1840}%
\special{pa 1000 160}%
\special{fp}%
\special{sh 1}%
\special{pa 1000 160}%
\special{pa 980 228}%
\special{pa 1000 214}%
\special{pa 1020 228}%
\special{pa 1000 160}%
\special{fp}%
%
\special{pn 20}%
\special{pa 1000 880}%
\special{pa 1240 880}%
\special{fp}%
%
\special{pn 20}%
\special{pa 1240 640}%
\special{pa 1480 640}%
\special{fp}%
%
\special{pn 20}%
\special{pa 1480 400}%
\special{pa 1960 400}%
\special{fp}%
%
\special{pn 20}%
\special{pa 1000 1120}%
\special{pa 760 1120}%
\special{fp}%
%
\special{pn 20}%
\special{pa 760 1360}%
\special{pa 520 1360}%
\special{fp}%
%
\special{pn 20}%
\special{pa 520 1600}%
\special{pa 40 1600}%
\special{fp}%
\put(10.6000,-2.2000){\makebox(0,0)[lb]{$y$}}%
\put(19.0000,-9.4000){\makebox(0,0)[lb]{$\psi$}}%
\put(9.4000,-8.8000){\makebox(0,0){$1$}}%
\put(9.4000,-6.4000){\makebox(0,0){$3$}}%
\put(9.4000,-4.0000){\makebox(0,0){$5$}}%
\put(10.6000,-11.2000){\makebox(0,0){$-1$}}%
\put(10.6000,-13.6000){\makebox(0,0){$-3$}}%
\put(10.6000,-16.0000){\makebox(0,0){$-5$}}%
\put(12.4000,-10.6000){\makebox(0,0){$2$}}%
\put(14.8000,-10.6000){\makebox(0,0){$4$}}%
\put(7.6000,-10.6000){\makebox(0,0){$-2$}}%
\put(5.2000,-10.6000){\makebox(0,0){$-4$}}%
\put(9.4000,-10.6000){\makebox(0,0){$0$}}%
\end{picture}%
\caption{Uniform quantizer $Q$ with $M=5$ (number of steps) and $\Delta=2\delta=2$ (step size).}
\label{fig:quantizerIO}
\end{figure}%
We here discuss the stability of the $\Delta\Sigma$ modulator model without linearization.

\subsection{Stability analysis in state space}
\label{subsec:stability:ss}

Let us first make the following assumptions:
\begin{ass}
\label{ass:q1}
The linearized model shown in Fig.~\ref{fig:linear-modulator} is internally stable.
That is, the filter $H(z)=[H_1(z), H_2(z)]$ satisfies \eqref{eq:param}.
\end{ass}
\begin{ass}
\label{ass:q2}
There exist real numbers $M>0$ and $\delta>0$ such that if $|\psi|\leq M+1$
then $|Q\psi - \psi|\leq \delta$.
\end{ass}

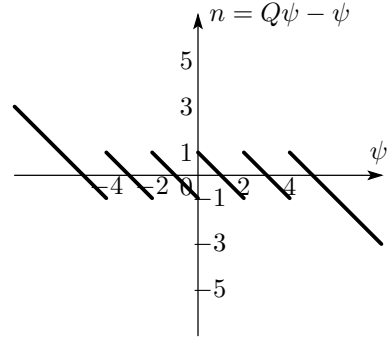
\begin{figure}[tbp]
\centering
\unitlength 0.1in
\begin{picture}( 19.2000, 17.9000)(  0.4000,-18.4000)
%
\special{pn 8}%
\special{pa 40 1000}%
\special{pa 1960 1000}%
\special{fp}%
\special{sh 1}%
\special{pa 1960 1000}%
\special{pa 1894 980}%
\special{pa 1908 1000}%
\special{pa 1894 1020}%
\special{pa 1960 1000}%
\special{fp}%
%
\special{pn 8}%
\special{pa 1000 1840}%
\special{pa 1000 160}%
\special{fp}%
\special{sh 1}%
\special{pa 1000 160}%
\special{pa 980 228}%
\special{pa 1000 214}%
\special{pa 1020 228}%
\special{pa 1000 160}%
\special{fp}%
\put(10.6000,-2.2000){\makebox(0,0)[lb]{$n=Q\psi-\psi$}}%
\put(19.0000,-9.4000){\makebox(0,0)[lb]{$\psi$}}%
\put(9.4000,-8.8000){\makebox(0,0){$1$}}%
\put(9.4000,-6.4000){\makebox(0,0){$3$}}%
\put(9.4000,-4.0000){\makebox(0,0){$5$}}%
\put(10.6000,-11.2000){\makebox(0,0){$-1$}}%
\put(10.6000,-13.6000){\makebox(0,0){$-3$}}%
\put(10.6000,-16.0000){\makebox(0,0){$-5$}}%
\put(12.4000,-10.6000){\makebox(0,0){$2$}}%
\put(14.8000,-10.6000){\makebox(0,0){$4$}}%
\put(7.6000,-10.6000){\makebox(0,0){$-2$}}%
\put(5.2000,-10.6000){\makebox(0,0){$-4$}}%
\put(9.4000,-10.6000){\makebox(0,0){$0$}}%
%
\special{pn 20}%
\special{pa 1000 880}%
\special{pa 1240 1120}%
\special{fp}%
%
\special{pn 20}%
\special{pa 1240 880}%
\special{pa 1480 1120}%
\special{fp}%
%
\special{pn 20}%
\special{pa 1480 880}%
\special{pa 1960 1360}%
\special{fp}%
%
\special{pn 20}%
\special{pa 760 880}%
\special{pa 1000 1120}%
\special{fp}%
%
\special{pn 20}%
\special{pa 760 1120}%
\special{pa 520 880}%
\special{fp}%
%
\special{pn 20}%
\special{pa 520 1120}%
\special{pa 40 640}%
\special{fp}%
\end{picture}%
\caption{Quantization error $n=Q\psi-\psi$ of the quantizer $Q$ in Fig.~\ref{fig:quantizerIO}.}
\label{fig:quantizererror}
\end{figure}

Note that the first assumption is necessary for the stability of the nonlinear system.
The second assumption considers general quantizers including uniform ones.
For example, the uniform quantizer shown in Fig.~\ref{fig:quantizerIO} 
has $M=5$ and $\delta=1$; see also Fig.~\ref{fig:quantizererror}.
For uniform quantizers, 
the number $\Delta=2\delta$ is called the \emph{step size} 
and the interval $[-M-1,M+1]$ is called
the \emph{no-overload input range}~\cite{SchTem}.
Under these assumptions, we have the following lemma:
\begin{lemma}
\label{lem:stability1}
Assume that Assumptions \ref{ass:q1} and \ref{ass:q2} hold.
If
$\psi(0)\leq M+1$
and if
$\|p\|_1\|u\|_\infty + \delta\|r\|_1 \leq M+1$,
then we have
\begin{equation}
\label{eq:lem:stability}
|n(k)|\leq\delta, \quad |\psi(k)|\leq M+1,\quad k=0,1,2,\dots,
\end{equation}
where $p$ and $r$ are respectively the impulse responses of $P$ and $R$,
and $\|\cdot\|_1$ and  $\|\cdot\|_\infty$ denote, respectively, the $\ell^1$ norm and
$\ell^\infty$ norm of sequences.
\end{lemma}
\begin{IEEEproof}
Since the filter $H=[H_1, H_2]$ satisfies \eqref{eq:param},
we have $\psi = Pu + Rn$ where $n:=Q\psi - \psi$.
Then, we have $\psi(k) = (p\ast u)(k) + (r\ast n)(k)$
for $k=0,1,2,\dots$.  
It follows that
\[
\begin{split}
|\psi(k)| 
&\leq |(p\ast u)(k)| + \sum_{i=1}^{k}|r(i)||n(k-i)|\\
&\leq \|p\ast u\|_\infty 
   + \left(\max_{0\leq i \leq k-1}|n(i)|\right)\sum_{i=1}^{k} |r(i)|.
\end{split}
\]
If $|\psi(0)|\leq M+1$, then by Assumption~\ref{ass:q2}, we have
$|n(0)|=|Q\psi(0)-\psi(0)|\leq \delta$, and hence
\[
\begin{split}
|\psi(1)|
&\leq\|p\ast u\|_\infty + \delta \sum_{i=1}^{k}|r(i)|\\
&\leq \|p\|_1\|u\|_\infty + \delta\|r\|_1\leq M+1.
\end{split}
\]
Again by Assumption~\ref{ass:q2}, we also have $|n(1)|\leq \delta$.
By induction on $k$, we deduce that 
$|\psi(k)|\leq M+1$ implies 
$|\psi(k+1)|\leq M+1$ and $|n(k+1)|\leq \delta$.
We thus have inequality \eqref{eq:lem:stability}.  
\end{IEEEproof}

This lemma gives a sufficient condition for the input $\psi$ 
of the quantizer $Q$ to be always in the no-overload range $[-M-1,M+1]$.
A $\Delta\Sigma$ modulator is conventionally said to be stable
if $\psi(k)\in [-M-1,M+1]$ for all $k\geq 0$~\cite{KenCar93,SchTem}.
However, since the modulator involves feedback, this does not
necessarily guarantee boundedness of all signals in the 
feedback loop.  
To show the boundedness, 
we introduce a state-space model of the $\Delta\Sigma$ modulator
for analyzing the stability of the feedback system.

First, invoke a minimal realization
of the filter $H(z)$ be $\{A_H, [B_1,B_2], C_H, [D_H,0]\}$,
as follows: 
\[
 \begin{split}
  H_1(z) &= C_H(zI-A_H)^{-1}B_1 + D_H,\\
  H_2(z) &= C_H(zI-A_H)^{-1}B_2.
 \end{split}
\]
Then a state-space model of the closed-loop 
system shown in Fig.~\ref{fig:delta-sigma} is given by the
following formulas: 
\begin{equation}
\begin{split}
x(k+1) &= A_{\cl}x(k) + B_uu(k) + B_nn(k),\\
n(k) &= (Q\psi-\psi)(k),\\
\psi(k) &= C_Hx(k)+D_Hu(k),\quad k=0,1,2,\dots,\\
A_{\cl}&:= A_H+B_2C_H,\\
B_u&:=B_1+B_2D_H,\quad
B_n:=B_2.
\label{eq:ss}
\end{split}
\end{equation}
The nonlinear effect of $Q$ is represented by the signal $n(k)$.

Consider the ideal state $x_\I(k)$, which is the state when there is no quantization,
that is, when $Q$ is identity (or $n\equiv 0$).
Define the state error $e:=x-x_\I$.
We then have the following theorem:
\begin{theorem}
\label{thm:q}
Suppose that the $\Delta\Sigma$ modulator shown in Fig.~\ref{fig:delta-sigma}
satisfies Assumptions \ref{ass:q1} and \ref{ass:q2}.
If
$\psi(0)\leq M+1$
and if
\begin{equation}
\|p\|_1\|u\|_\infty + \delta \|r\|_1 \leq M+1,
\label{eq:theorem1}
\end{equation}
then there exists a bounded, 
real and monotone increasing sequence $\{\beta_k\}$
such that
\begin{equation}
\label{eq:thm:q}
|e(k)|\leq \beta_k,\quad k=0,1,2,\dots,
\end{equation}
where $|e(k)|$ denotes the Euclidean norm of vector $e(k)$.
\end{theorem}
\begin{IEEEproof}
By the state-space representation \eqref{eq:ss}, we have
\[
\begin{split}
x(k) &= A^k_{\cl}x(0) + \sum_{i=0}^{k-1}A^i_{\cl}B_uu(k-i) 
         + \sum_{i=0}^{k-1}A^i_{\cl}B_nn(k-i)\\
     &= x_\I(k) + \sum_{i=0}^{k-1}A^i_{\cl}B_nn(k-i).
\end{split}
\]
From this, we obtain
\[
  e(k) = x(k)-x_\I(k) = \sum_{i=0}^{k-1} A_{\cl}^iB_nn(k-i).
\]
By the triangle inequality, we have
\[
  |e(k)| \leq \sum_{i=0}^{k-1} \|A^i_{\cl}B_n\|\cdot\lvert n(k-i) \rvert.
\]
From Lemma~\ref{lem:stability1},  
we have $|n(k)|\leq\delta$ for all $k\geq 0$.
Put
\[
\beta_k := \delta\sum_{i=0}^{k-1} \|A^i_{\cl}B_n\|.
\]
Since matrix $A_{\cl}$ is stable by Assumption~\ref{ass:q1}, 
the sequence $\{\beta_k\}_{k\geq 0}$ is bounded and monotone increasing,
and we have
$|e(k)| \leq \beta_k$ for all $k=0,1,2,\dots$.
\end{IEEEproof}

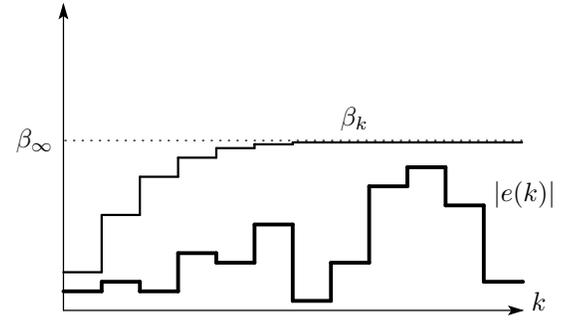
\begin{figure}[tbp]
\centering
\unitlength 0.1in
\begin{picture}( 32.3000, 16.0500)( -3.8000,-18.0500)
%
\special{pn 8}%
\special{pa 400 1800}%
\special{pa 400 200}%
\special{fp}%
\special{sh 1}%
\special{pa 400 200}%
\special{pa 380 268}%
\special{pa 400 254}%
\special{pa 420 268}%
\special{pa 400 200}%
\special{fp}%
%
\special{pn 13}%
\special{pa 400 1600}%
\special{pa 600 1600}%
\special{fp}%
%
\special{pn 13}%
\special{pa 600 1600}%
\special{pa 600 1300}%
\special{fp}%
\special{pa 600 1300}%
\special{pa 800 1300}%
\special{fp}%
\special{pa 800 1300}%
\special{pa 800 1100}%
\special{fp}%
\special{pa 800 1100}%
\special{pa 1000 1100}%
\special{fp}%
\special{pa 1000 1100}%
\special{pa 1000 1000}%
\special{fp}%
\special{pa 1000 1000}%
\special{pa 1200 1000}%
\special{fp}%
\special{pa 1200 1000}%
\special{pa 1200 950}%
\special{fp}%
\special{pa 1200 950}%
\special{pa 1400 950}%
\special{fp}%
\special{pa 1400 930}%
\special{pa 1400 930}%
\special{fp}%
\special{pa 1400 950}%
\special{pa 1400 930}%
\special{fp}%
\special{pa 1400 930}%
\special{pa 1600 930}%
\special{fp}%
\special{pa 1600 930}%
\special{pa 1600 920}%
\special{fp}%
\special{pa 1600 920}%
\special{pa 2800 920}%
\special{fp}%
%
\special{pn 8}%
\special{pa 400 1800}%
\special{pa 2800 1800}%
\special{fp}%
\special{sh 1}%
\special{pa 2800 1800}%
\special{pa 2734 1780}%
\special{pa 2748 1800}%
\special{pa 2734 1820}%
\special{pa 2800 1800}%
\special{fp}%
%
\special{pn 20}%
\special{pa 400 1700}%
\special{pa 600 1700}%
\special{fp}%
\special{pa 600 1700}%
\special{pa 600 1650}%
\special{fp}%
\special{pa 600 1650}%
\special{pa 800 1650}%
\special{fp}%
\special{pa 800 1650}%
\special{pa 800 1700}%
\special{fp}%
%
\special{pn 20}%
\special{pa 800 1700}%
\special{pa 1000 1700}%
\special{fp}%
\special{pa 1000 1700}%
\special{pa 1000 1500}%
\special{fp}%
\special{pa 1000 1500}%
\special{pa 1200 1500}%
\special{fp}%
\special{pa 1200 1500}%
\special{pa 1200 1550}%
\special{fp}%
\special{pa 1200 1550}%
\special{pa 1400 1550}%
\special{fp}%
\special{pa 1400 1550}%
\special{pa 1400 1350}%
\special{fp}%
\special{pa 1400 1350}%
\special{pa 1600 1350}%
\special{fp}%
\special{pa 1600 1350}%
\special{pa 1600 1750}%
\special{fp}%
\special{pa 1600 1750}%
\special{pa 1800 1750}%
\special{fp}%
\special{pa 1800 1750}%
\special{pa 1800 1550}%
\special{fp}%
\special{pa 1800 1550}%
\special{pa 2000 1550}%
\special{fp}%
\special{pa 2000 1550}%
\special{pa 2000 1150}%
\special{fp}%
\special{pa 2000 1150}%
\special{pa 2200 1150}%
\special{fp}%
\special{pa 2200 1150}%
\special{pa 2200 1050}%
\special{fp}%
\special{pa 2200 1050}%
\special{pa 2400 1050}%
\special{fp}%
\special{pa 2400 1050}%
\special{pa 2400 1250}%
\special{fp}%
\special{pa 2400 1250}%
\special{pa 2600 1250}%
\special{fp}%
\special{pa 2600 1250}%
\special{pa 2600 1650}%
\special{fp}%
\special{pa 2600 1650}%
\special{pa 2800 1650}%
\special{fp}%
\special{pa 2800 1650}%
\special{pa 2800 1650}%
\special{fp}%
\put(28.5000,-18.0000){\makebox(0,0)[lb]{$k$}}%
\put(18.5000,-8.6000){\makebox(0,0)[lb]{$\beta_k$}}%
\put(26.5000,-12.6000){\makebox(0,0)[lb]{$|e(k)|$}}%
%
\special{pn 8}%
\special{pa 2800 910}%
\special{pa 400 910}%
\special{dt 0.045}%
\put(2.5000,-9.1000){\makebox(0,0){$\beta_\infty$}}%
\end{picture}%
\caption{Boundedness of quantization error $|e(k)|$, where $\beta_\infty$
is the limiting value of $\{\beta_k\}$.}
\label{fig:stability}
\end{figure}

Stability condition \eqref{eq:theorem1} depends on
the maximum amplitude of the input $u$.
This is different from stability condition \eqref{eq:param}
for the linearized model that is independent of $u$.
The difference is due to the nonlinearity (in particular, saturation)
in the quantizer $Q$.
Therefore, one should limit the level of inputs before it is quantized.
See also the example in Section~\ref{subsec:example_lp}.
From Theorem~\ref{thm:q}, it follows that when a $\Delta\Sigma$ modulator 
satisfies the condition in Theorem~\ref{thm:q},
the error $|e(k)|$ in the state space is bounded as shown in
Fig.~\ref{fig:stability}.
As a result, the state $x(k)$ is also bounded, and
we can conclude that the system is stable in a weak sense
(i.e., bounded but not guaranteed to converge to zero).
By Theorem~\ref{thm:q}, we derive a generalization of the stability condition given in~\cite{KenCar93}
as the following corollary:
\begin{cor}
\label{co:q}
Suppose that the $\Delta\Sigma$ modulator shown in Fig.~\ref{fig:delta-sigma}
satisfies Assumptions \ref{ass:q1} and \ref{ass:q2}.
Define the noise-to-state transfer function $G(z)$ by 
$G(z) = (zI-A_{\cl})^{-1}B_n$,
and its impulse response by $g$.
If $\psi(0)\leq M+1$
and if inequality \eqref{eq:theorem1} holds,
then we have
$\|e\|_\infty \leq \delta \|g\|_1$.
\end{cor}
\begin{IEEEproof}
By Theorem~\ref{thm:q}, we have
\[
 |e(k)| \leq \lim_{k\rightarrow\infty}\beta_k = \delta\sum_{i=0}^\infty \|A_{\cl}^iB_n\| = \delta\|g\|_1,
\]
for all $k=0,1,2,\dots$.
\end{IEEEproof}

\subsection{Stability condition by an $H^\infty$ norm inequality}
\label{subsec:stability-norm}

Assume that $\|p\|_1=1$.
Then, we can rewrite condition \eqref{eq:theorem1} in Theorem~\ref{thm:q} as
\begin{equation}
\label{eq:l1}
\|r\|_1 \leq \frac{1}{\delta} (M+1-\|u\|_\infty).
\end{equation}
By \eqref{eq:FIR}, we have
$\|1+r\|_1=1+\sum_{k=1}^N|\alpha_k| = 1+\|r\|_1$,
and we can show that \eqref{eq:l1} is 
equivalent to the condition given in~\cite{KenCar93,SchTem}:
\begin{equation}
\|1 + r\|_1 \leq \frac{1}{\delta} (M+1+\delta-\|u\|_\infty).
\label{eq:l12}
\end{equation}
Let $N$ be the order of $R(z)$.
Then by the following inequality (see~\cite[Theorem 4.3.1]{DahDiaz}),
\[
\|1+r\|_1 \leq (2N+1)\|1+R\|_\infty,
\]
we have 
a sufficient condition for \eqref{eq:l12}:
\begin{equation}
\label{eq:stab_cond22}
\begin{split}
\|\NTF\|_\infty&=\|1+R\|_\infty\\
 &\leq \frac{1}{(2N+1)\delta}(M+1+\delta-\|u\|_\infty).
\end{split}
\end{equation}

For the stability of binary $\Delta\Sigma$ modulators, the following criterion%
\footnote{Note that this is 
neither sufficient nor necessary for stability.},
called the Lee criterion,
is widely used~\cite{ChaNadLeeSod90,SchTem}:
\begin{equation}
\|\NTF\|_\infty=\|1+R\|_\infty < 1.5.
\label{eq:Lee}
\end{equation}

From conditions \eqref{eq:stab_cond22} and \eqref{eq:Lee},
attenuation of the $H^\infty$ norm of $\NTF=1+R$ improves the stability.  
Therefore, we add the following stability constraints to the design of modulators:
\[
\|\NTF\|_\infty=\|1+R\|_\infty < \gamma_0,
\]
where $\gamma_0>0$ is a constant (e.g., $\gamma_0=1.5$ for the Lee criterion).
Assuming that $R(z)$ is the FIR filter defined
by \eqref{eq:FIR}
and also that its state-space matrices are given in \eqref{eq:R-ABC},
the above inequality is also reducible to an LMI via the KYP lemma, 
also known as the bounded real lemma~\cite{BoyGhaFerBal,YamAndNagKoy03}:
\begin{lemma}
\label{lem:hinf}
The inequality $\|\NTF\|_\infty<\gamma_0$ holds if and only if
there exists a symmetric matrix $Z>0$ such that
\[
 \begin{bmatrix}
  A\tr Z A - Z & A\tr Z B & C(\alpha)\tr\\ B\tr Z A & B\tr ZB 
   - \gamma_0^2 & 1\\ C(\alpha) & 1 & -1
 \end{bmatrix} <0.
\]
\end{lemma}
\begin{IEEEproof}
The equivalence is a direct consequence of the KYP lemma 
(aka, bounded real lemma)~\cite[Sec.~2.7]{BoyGhaFerBal}
and the Schur complement~\cite[Sec.~2.1]{BoyGhaFerBal}.
\end{IEEEproof}

\section{Cascade of Error Feedback for High-order modulators}
\label{sec:cascade}

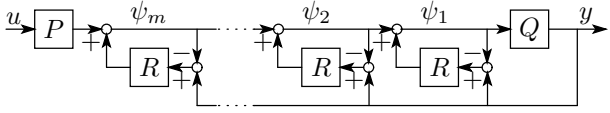
\begin{figure}[tbp]
\centering
\unitlength 0.1in
\begin{picture}( 31.4000,  5.9500)(  4.0000, -6.9000)
%
\special{pn 8}%
\special{pa 400 290}%
\special{pa 550 290}%
\special{fp}%
\special{sh 1}%
\special{pa 550 290}%
\special{pa 484 270}%
\special{pa 498 290}%
\special{pa 484 310}%
\special{pa 550 290}%
\special{fp}%
%
\special{pn 8}%
\special{pa 550 190}%
\special{pa 750 190}%
\special{pa 750 390}%
\special{pa 550 390}%
\special{pa 550 190}%
\special{fp}%
%
\special{pn 8}%
\special{pa 750 290}%
\special{pa 900 290}%
\special{fp}%
\special{sh 1}%
\special{pa 900 290}%
\special{pa 834 270}%
\special{pa 848 290}%
\special{pa 834 310}%
\special{pa 900 290}%
\special{fp}%
%
\special{pn 8}%
\special{ar 926 290 26 26  0.0000000 6.2831853}%
%
\special{pn 8}%
\special{pa 950 290}%
\special{pa 1500 290}%
\special{fp}%
%
\special{pn 8}%
\special{pa 926 490}%
\special{pa 926 316}%
\special{fp}%
\special{sh 1}%
\special{pa 926 316}%
\special{pa 906 382}%
\special{pa 926 368}%
\special{pa 946 382}%
\special{pa 926 316}%
\special{fp}%
%
\special{pn 8}%
\special{pa 1050 390}%
\special{pa 1250 390}%
\special{pa 1250 590}%
\special{pa 1050 590}%
\special{pa 1050 390}%
\special{fp}%
%
\special{pn 8}%
\special{pa 1050 490}%
\special{pa 926 490}%
\special{fp}%
%
\special{pn 8}%
\special{pa 1376 490}%
\special{pa 1250 490}%
\special{fp}%
\special{sh 1}%
\special{pa 1250 490}%
\special{pa 1318 510}%
\special{pa 1304 490}%
\special{pa 1318 470}%
\special{pa 1250 490}%
\special{fp}%
%
\special{pn 8}%
\special{ar 1400 490 26 26  0.0000000 6.2831853}%
%
\special{pn 8}%
\special{pa 1400 290}%
\special{pa 1400 466}%
\special{fp}%
\special{sh 1}%
\special{pa 1400 466}%
\special{pa 1420 398}%
\special{pa 1400 412}%
\special{pa 1380 398}%
\special{pa 1400 466}%
\special{fp}%
%
\special{pn 8}%
\special{pa 1700 290}%
\special{pa 1800 290}%
\special{fp}%
\special{sh 1}%
\special{pa 1800 290}%
\special{pa 1734 270}%
\special{pa 1748 290}%
\special{pa 1734 310}%
\special{pa 1800 290}%
\special{fp}%
%
\special{pn 8}%
\special{ar 1826 290 26 26  0.0000000 6.2831853}%
%
\special{pn 8}%
\special{pa 1822 490}%
\special{pa 1822 316}%
\special{fp}%
\special{sh 1}%
\special{pa 1822 316}%
\special{pa 1802 382}%
\special{pa 1822 368}%
\special{pa 1842 382}%
\special{pa 1822 316}%
\special{fp}%
%
\special{pn 8}%
\special{pa 1948 390}%
\special{pa 2148 390}%
\special{pa 2148 590}%
\special{pa 1948 590}%
\special{pa 1948 390}%
\special{fp}%
%
\special{pn 8}%
\special{pa 1948 490}%
\special{pa 1822 490}%
\special{fp}%
%
\special{pn 8}%
\special{pa 2272 490}%
\special{pa 2148 490}%
\special{fp}%
\special{sh 1}%
\special{pa 2148 490}%
\special{pa 2216 510}%
\special{pa 2202 490}%
\special{pa 2216 470}%
\special{pa 2148 490}%
\special{fp}%
%
\special{pn 8}%
\special{ar 2298 490 26 26  0.0000000 6.2831853}%
%
\special{pn 8}%
\special{pa 2298 290}%
\special{pa 2298 466}%
\special{fp}%
\special{sh 1}%
\special{pa 2298 466}%
\special{pa 2318 398}%
\special{pa 2298 412}%
\special{pa 2278 398}%
\special{pa 2298 466}%
\special{fp}%
%
\special{pn 8}%
\special{pa 2442 490}%
\special{pa 2442 316}%
\special{fp}%
\special{sh 1}%
\special{pa 2442 316}%
\special{pa 2422 382}%
\special{pa 2442 368}%
\special{pa 2462 382}%
\special{pa 2442 316}%
\special{fp}%
%
\special{pn 8}%
\special{pa 2568 390}%
\special{pa 2768 390}%
\special{pa 2768 590}%
\special{pa 2568 590}%
\special{pa 2568 390}%
\special{fp}%
%
\special{pn 8}%
\special{pa 2568 490}%
\special{pa 2442 490}%
\special{fp}%
%
\special{pn 8}%
\special{pa 2892 490}%
\special{pa 2768 490}%
\special{fp}%
\special{sh 1}%
\special{pa 2768 490}%
\special{pa 2836 510}%
\special{pa 2822 490}%
\special{pa 2836 470}%
\special{pa 2768 490}%
\special{fp}%
%
\special{pn 8}%
\special{ar 2918 490 26 26  0.0000000 6.2831853}%
%
\special{pn 8}%
\special{pa 2918 290}%
\special{pa 2918 466}%
\special{fp}%
\special{sh 1}%
\special{pa 2918 466}%
\special{pa 2938 398}%
\special{pa 2918 412}%
\special{pa 2898 398}%
\special{pa 2918 466}%
\special{fp}%
%
\special{pn 8}%
\special{ar 2446 290 26 26  0.0000000 6.2831853}%
%
\special{pn 8}%
\special{pa 1850 290}%
\special{pa 2426 290}%
\special{fp}%
\special{sh 1}%
\special{pa 2426 290}%
\special{pa 2358 270}%
\special{pa 2372 290}%
\special{pa 2358 310}%
\special{pa 2426 290}%
\special{fp}%
%
\special{pn 8}%
\special{pa 2466 290}%
\special{pa 3040 290}%
\special{fp}%
\special{sh 1}%
\special{pa 3040 290}%
\special{pa 2974 270}%
\special{pa 2988 290}%
\special{pa 2974 310}%
\special{pa 3040 290}%
\special{fp}%
%
\special{pn 8}%
\special{pa 3040 190}%
\special{pa 3240 190}%
\special{pa 3240 390}%
\special{pa 3040 390}%
\special{pa 3040 190}%
\special{fp}%
%
\special{pn 8}%
\special{pa 3240 290}%
\special{pa 3540 290}%
\special{fp}%
\special{sh 1}%
\special{pa 3540 290}%
\special{pa 3474 270}%
\special{pa 3488 290}%
\special{pa 3474 310}%
\special{pa 3540 290}%
\special{fp}%
%
\special{pn 8}%
\special{pa 3390 290}%
\special{pa 3390 690}%
\special{fp}%
\special{pa 3390 690}%
\special{pa 1700 690}%
\special{fp}%
%
\special{pn 8}%
\special{pa 1500 690}%
\special{pa 1400 690}%
\special{fp}%
%
\special{pn 8}%
\special{pa 1400 690}%
\special{pa 1400 516}%
\special{fp}%
\special{sh 1}%
\special{pa 1400 516}%
\special{pa 1380 582}%
\special{pa 1400 568}%
\special{pa 1420 582}%
\special{pa 1400 516}%
\special{fp}%
%
\special{pn 8}%
\special{pa 2300 690}%
\special{pa 2300 516}%
\special{fp}%
\special{sh 1}%
\special{pa 2300 516}%
\special{pa 2280 582}%
\special{pa 2300 568}%
\special{pa 2320 582}%
\special{pa 2300 516}%
\special{fp}%
%
\special{pn 8}%
\special{pa 2920 690}%
\special{pa 2920 516}%
\special{fp}%
\special{sh 1}%
\special{pa 2920 516}%
\special{pa 2900 582}%
\special{pa 2920 568}%
\special{pa 2940 582}%
\special{pa 2920 516}%
\special{fp}%
%
\special{pn 8}%
\special{pa 1700 690}%
\special{pa 1500 690}%
\special{dt 0.045}%
%
\special{pn 8}%
\special{pa 1700 290}%
\special{pa 1500 290}%
\special{dt 0.045}%
\put(9.0000,-3.1500){\makebox(0,0)[rt]{$+$}}%
\put(18.0000,-3.1500){\makebox(0,0)[rt]{$+$}}%
\put(24.2000,-3.1500){\makebox(0,0)[rt]{$+$}}%
\put(28.9500,-5.1500){\makebox(0,0)[rt]{$+$}}%
\put(22.7500,-5.1500){\makebox(0,0)[rt]{$+$}}%
\put(13.7500,-5.1500){\makebox(0,0)[rt]{$+$}}%
\put(13.7500,-4.6500){\makebox(0,0)[rb]{$-$}}%
\put(22.7500,-4.6500){\makebox(0,0)[rb]{$-$}}%
\put(28.9500,-4.6500){\makebox(0,0)[rb]{$-$}}%
\put(4.0000,-2.6500){\makebox(0,0)[lb]{$u$}}%
\put(34.0000,-2.6500){\makebox(0,0)[lb]{$y$}}%
\put(31.4000,-2.9000){\makebox(0,0){$Q$}}%
\put(26.7000,-4.9000){\makebox(0,0){$R$}}%
\put(20.5000,-4.9000){\makebox(0,0){$R$}}%
\put(11.5000,-4.9000){\makebox(0,0){$R$}}%
\put(6.5000,-2.9000){\makebox(0,0){$P$}}%
\put(10.5000,-2.6500){\makebox(0,0)[lb]{$\psi_m$}}%
\put(19.5000,-2.6500){\makebox(0,0)[lb]{$\psi_2$}}%
\put(25.7000,-2.6500){\makebox(0,0)[lb]{$\psi_1$}}%
\end{picture}%
\caption{Cascade of Error Feedback}
\label{fig:cascade}
\end{figure}

To design a high-order modulator,
we can use the cascade construction of the error feedback modulators 
in Fig.~\ref{fig:mod2}.
The proposed cascade structure is shown in 
Fig.~\ref{fig:cascade}.
By using this structure,
we have
$\STF(z)=P(z)$ and 
\[
\NTF(z) = \bigl(1+R(z)\bigr)^m,
\]
where $m$ denotes the number of filters $R(z)$.
This can be proved by the following equations:
\[
 \begin{split}
  \psi_m &= Pu + R(y-\psi_m),\\
  y-\psi_{k} &= (1+R)(y-\psi_{k-1}),\quad k=m,m-1,\dots,2,\\
  y-\psi_1 &= n.
 \end{split}
\]
If $R(z)\in{\mathcal{S}}'$, then the 
linearized feedback system
is stable.
An advantage of this structure is that
the number of taps of $R(z)$ can be reduced,
and hence the implementation is much easier
than a filter with a large number of taps.
This structure can be applied to
$\Delta\Sigma$ DA converters.

To satisfy the stability condition
$\|\NTF\|_\infty<\gamma_0$,
the filter $R(z)$ is designed to limit
$\|1+R\|_\infty < \sqrt[m]{\gamma_0}$.
If this is satisfied, we have
$\|\NTF\|_\infty \leq \|1+R\|_\infty^m < \gamma_0$,
by the sub-multiplicative property of the $H^\infty$ norm~\cite{DahDiaz}.

\section{Design Examples}
\label{sec:examples}

In this section, we show two design examples of lowpass
and bandpass $\Delta\Sigma$ modulators by the proposed method.

\subsection{Lowpass modulator}
\label{subsec:example_lp}

We here show a design example of a high-order
lowpass modulator with the cascade structure shown in Fig.~\ref{fig:cascade}.
We set $P(z)=1$, that is, $\STF(z)=1$, and $R(z)$ be an FIR filter with 32 taps.
The cutoff frequency $\Omega$ is set to be $\pi/32$.
The FIR filter $R(z)$ is designed to minimize $\Vw(\NTF,[0,\Omega])$
defined in \eqref{eq:Vw} and the coefficients are obtained
by  the LMI in Theorem~\ref{thm:gkyp},
with the stability condition $\|\NTF\|_\infty < 1.5$,
which is also described by an LMI in Lemma~\ref{lem:hinf}. 
The number $m$ of cascades is 2,
that is, the order of the modulator is $32\times 2=64$.
We also design a modulator by
the NTF zero optimization~\cite{Sch93,SchTem}
that minimizes the average $\Va(\NTF,[0,\Omega])$ defined in \eqref{eq:Va}.
This modulator is designed by 
the MATLAB function \verb=synthesizeNTF=
in the Delta-Sigma Toolbox~\cite{SchTem,DST},
where the order of $\NTF$ is 4,
the over sampling ratio $\OSR$ is 32,
and the stability condition $\|\NTF\|_\infty<1.5$.

\begin{figure}[t]
\centering
\includegraphics[width=\linewidth]{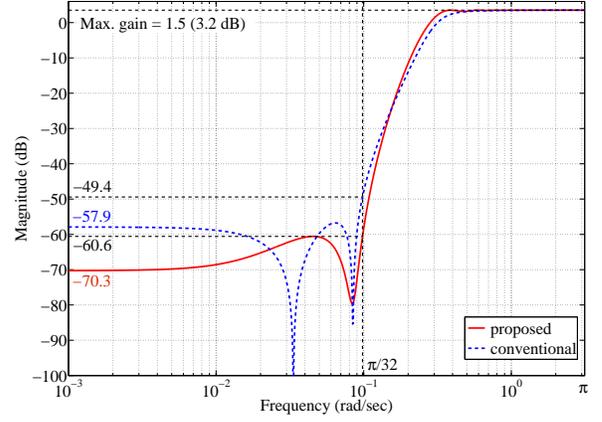}
\caption{NTF's: proposed (solid) and the NTF zero optimization (dash).}
\label{fig:NTF}
\end{figure}

Fig.~\ref{fig:NTF} shows the frequency responses
of the proposed modulator and that by optimizing the NTF zeros.
By this figure, we see that the magnitude of the proposed NTF is uniformly
attenuated over $[0,\pi/32]$
while the conventional one shows peaks in this band.
The difference between the two maximal magnitudes 
at the frequency $\omega=\pi/32$ is approximately $11.2$ (dB),
and the difference at low frequencies is about $12.4$ (dB).

\begin{figure}[t]
\centering
\includegraphics[width=\linewidth]{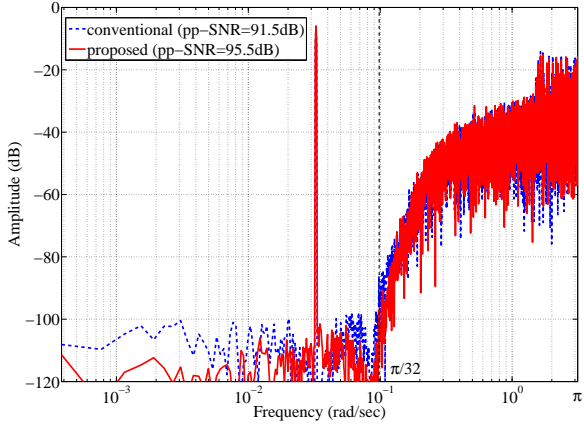}
\caption{The spectrum of the output of the $\Delta\Sigma$ modulators:
   proposed (solid) and conventional (dash), pp-SNR denotes the peak-to-peak SNR.}
\label{fig:freqresp}
\end{figure}

Then we run a simulation to evaluate the obtained modulators.
We used MATLAB functions \verb=simulateDSM= and \verb=simulateSNR= in
the Delta-Sigma Toolbox.
Fig.~\ref{fig:freqresp} shows the spectrum of the output
when the input is the sinusoidal wave with frequency 0.0325 (rad/sec)
and amplitude 0.5.
We assume a uniform quantizer with $M=1$ and $\delta=1/2$ (see Assumption~\ref{ass:q2}).
We observe that the quantization noise is well attenuated in both cases.
Note that the frequency 0.0325 (rad/sec) is taken around the first
notch of the conventional NTF gain (see Fig.~\ref{fig:NTF}).
The notch frequency is expected to give 
much better performance to the conventional modulator
than the proposed modulator.
However, the simulation shows this does not necessarily hold.
In fact, the peak-to-peak SNR, $\SNR$ defined in \eqref{eq:SNRworst},
of our modulator is 95.5 (dB), while
that of the conventional modulator is 91.5 (dB).
That is, our design is superior to the conventional one in $\SNR$ by approximately 4.0 (dB).

\begin{figure}[t]
\centering
\includegraphics[width=\linewidth]{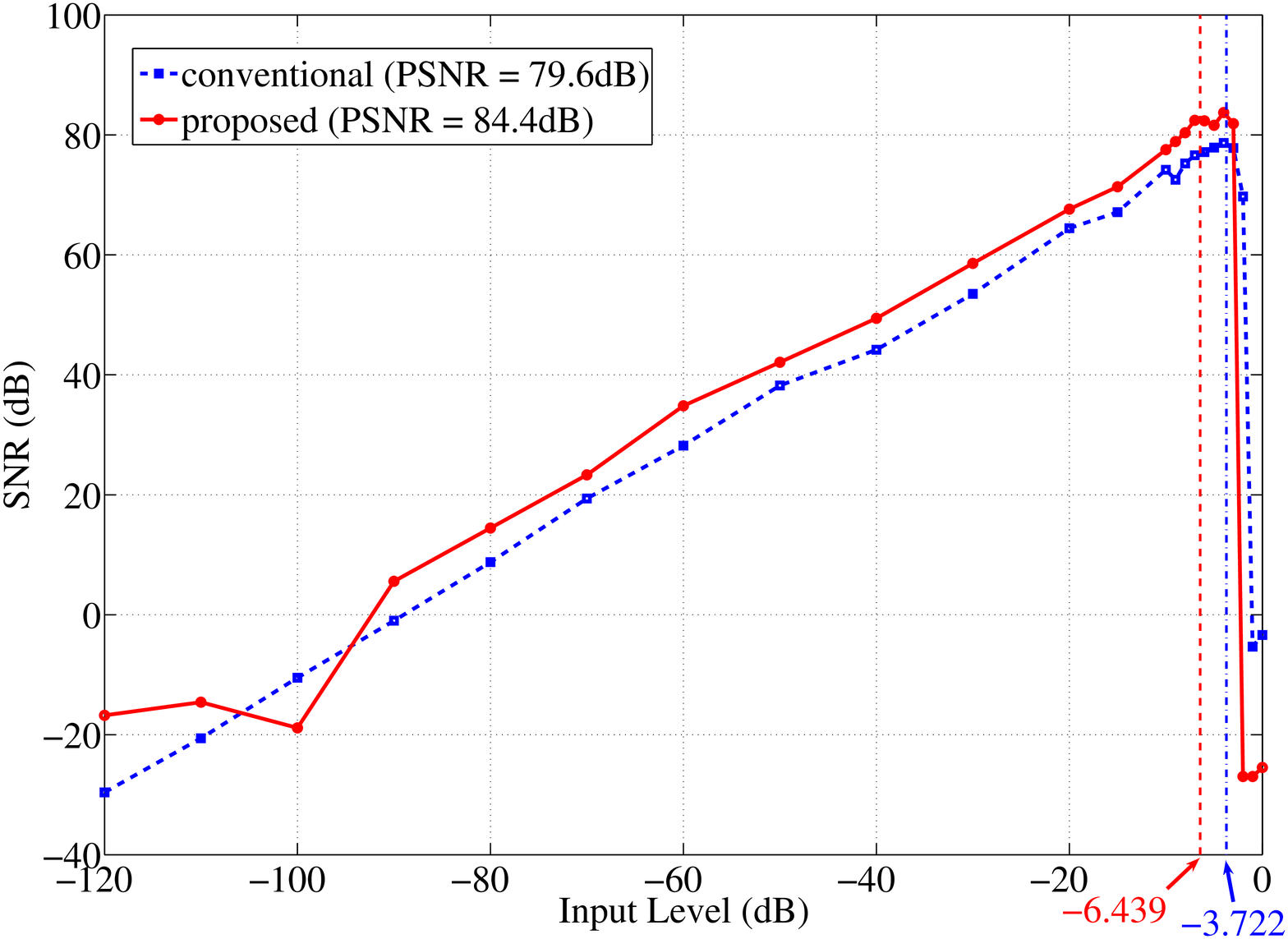}
\caption{The SNR versus the amplitude of the input: 
  proposed (solid) and conventional (dash).
  -6.439 is the stability bound for the proposed modulator, and
  -3.722 is for the conventional modulator.}
\label{fig:DR_SNR}
\end{figure}

\begin{figure}[t]
\centering
\includegraphics[width=\linewidth]{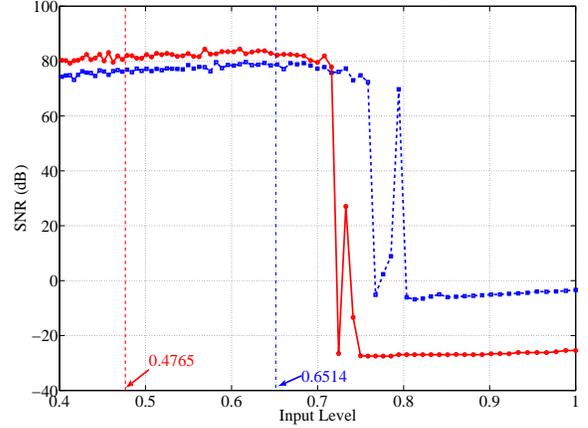}
\caption{Enlarged plot of Fig.~\ref{fig:DR_SNR}
with linear scale for input levels.}
\label{fig:DR_SNR_w}
\end{figure}

Fig.~\ref{fig:DR_SNR} shows the SNR,
the ratio of the signal power to the quantization noise power (SQNR),
of the modulators as a function of the amplitude
of the input sinusoidal wave with the frequency 0.0325 (rad/sec).
For almost all amplitudes,
the proposed modulator shows better performance than the conventional one,
in particular, the difference of the peak SNR, or the maximum SNR is about 
4.8 (dB).
The figure also shows the stability bounds estimated by 
inequality \eqref{eq:theorem1} in Theorem~\ref{thm:q}.
That is, the bound for the conventional modulator is given by
$M+1-\delta\|r\|_1\approx 0.6514$ (-3.722 dB),
and that for the proposed modulator is
$M+1-\delta\|r\|_1\approx 0.4765$ (-6.439 dB).
The degradation of the SNR for high input levels
is due
to saturation in the quantizer that leads to
instability in
the modulator.
We can say that if the input level is limited to the stability bound,
the degradation is avoidable.
We note that the conventional modulator can accept higher level of inputs.
To see the difference more precisely,
we show an enlarged plot in Fig.~\ref{fig:DR_SNR_w}.
The difference however does not matter if the inputs are limited to
the pre-estimated bound by Theorem~\ref{thm:q}.
These simulation results show that the proposed min-max
(or worst-case) design gives a better SNR as mentioned
in Section~\ref{subsec:worstcase}.
We summarize the results in Table~\ref{table:comparison}.
\begin{table}[t]
\centering
\caption{Comparison in Figs.~\ref{fig:NTF}--\ref{fig:DR_SNR}.}
\label{table:comparison}
\begin{tabular}{|c|c|c|c|c|}\hline
 & max NTF (dB)& $\SNR$ (dB) & peak SNR (dB)\\\hline
Conventional & -49.4 & 91.5 & 79.6\\
Proposed & -60.6 & 99.5 & 84.4\\\hline
Improvement & 11.2 & 4.0 & 4.8\\\hline
\end{tabular}
\end{table}

\subsection{Bandpass modulator}
\label{subsec:example_bp}

We next show a design example of a bandpass modulator.
We set $P(z)=1$, and $R(z)$ be an FIR filter with 32 taps.
The center frequency $\omega_0$ is set to be $\pi/2$,
and the bandwidth parameter $\Omega$ is $\pi/16$.
The FIR filter $R(z)$ is designed by using the LMI
in Theorem~\ref{thm:gkyp_bp},
with the stability condition $\|\NTF\|_\infty < 1.5$.
We design two modulators, with zeros at $\omega_0=\pm \pi/2$
and without assignment of zeros there.
We also design a modulator by
the NTF zero optimization~\cite{Sch93,SchTem},
designed by the MATLAB function \verb=synthesizeNTF=
in the Delta-Sigma Toolbox,
with the order of $\NTF$ is 6,
the over sampling ratio $\OSR$ is 16,
the center frequency $f_0=1/4$,
and $\|\NTF\|_\infty<1.5$.

\begin{figure}[t]
\centering
\includegraphics[width=\linewidth]{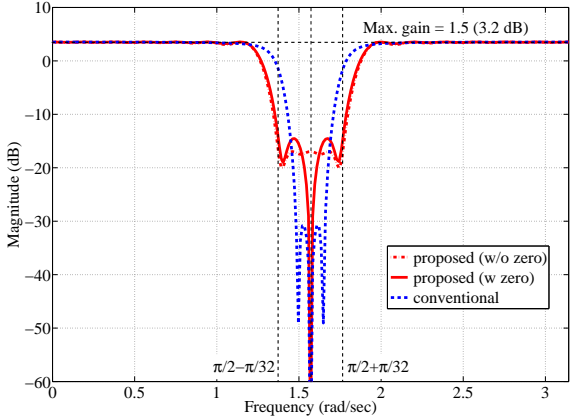}
\caption{Bandpass NTF's: proposed with zeros at $\omega_0=\pm \pi/2$ (solid),
proposed without assignment of zeros (dash-dots) and the NTF zero optimization (dash).}
\label{fig:NTF_bp}
\end{figure}%
\begin{figure}[t]
\centering
\includegraphics[width=\linewidth]{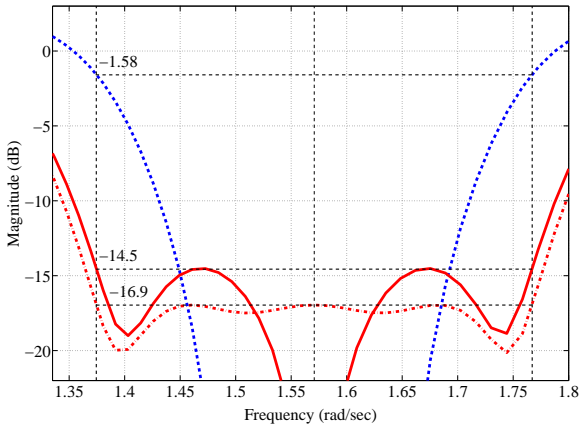}
\caption{Enlarged view of bandpass NTF's in Fig.~\ref{fig:NTF_bp}.}
\label{fig:NTF_bp_e}
\end{figure}%

Fig.~\ref{fig:NTF_bp} shows the frequency responses
of the two proposed modulators and that by optimizing the NTF zeros.
We can see that the proposed modulator without assignment of zeros
shows the smallest magnitude over the band $[\pi/2-\pi/16,\pi/2+\pi/16]$,
and that of the proposed modulator with a zero at $\pi/2$ is slightly larger.
To see these precisely, enlarged figure of Fig.~\ref{fig:NTF_bp}
around the center frequency is shown in Fig.~\ref{fig:NTF_bp_e}.
By this figure, the magnitudes of the proposed NTF's are uniformly
attenuated over the band,
while the conventional one shows a peak on the edges of the band.
The differences between the magnitudes of the proposed NTF's and that of the conventional one
are about 12.9 (dB) and 15.3 (dB).

\begin{figure}[t]
\centering
\includegraphics[width=\linewidth]{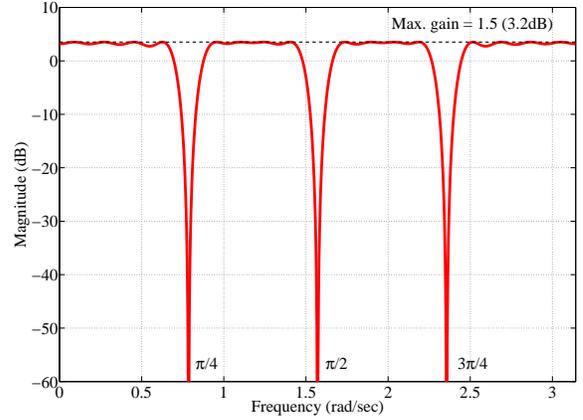}
\caption{Multi-band bandpass NTF designed by Theorem~\ref{thm:gkyp_mb}
with zeros at $\omega_1=\pi/4$, $\omega_2=\pi/2$, and $\omega_3=3\pi/4$.}
\label{fig:NTF_mb}
\end{figure}

Finally, we give an example of a multi-band modulator
proposed in Section~\ref{subsec:gkyp_bp}.
We set $P(z)=1$, and $R(z)$ be an FIR filter with 32 taps.
The center frequencies are set by $\omega_1=\pi/4$,
$\omega_2=\pi/2$, and $\omega_3=3\pi/4$.
The bandwidth parameter is $\Omega_l=\pi/16$,
$l=1,2,3$.
We also impose the infinity norm condition $\|\NTF\|<1.5$ and place
zeros at $\omega_1$, $\omega_2$, and $\omega_3$.
Fig.~\ref{fig:NTF_mb} shows the magnitude frequency response of
the NTF designed via Theorem~\ref{thm:gkyp_mb}.
The figure shows that our design method works well.

\section{Conclusion}
\label{sec:conclusion}

We have proposed a min-max design method of $\Delta\Sigma$
modulators.
First we have characterized all stabilizing loop-filters 
for a linearized model.  
Then, based on this result, 
we have formulated our problem of 
noise shaping in the frequency domain 
as a min-max optimization.  
It is seen that the proposed min-max design
has an advantage in improving SNR.  

The proposed design problem is reduced to an LMI optimization,
using the generalized KYP lemma, and this has 
a computational advantage.  The 
assignment of NTF zeros can be taken care of by an LME.  
We have given a stability analysis of 
the $\Delta\Sigma$ modulator model without linearization
and derived an $H^\infty$-norm condition for stability,
which is also described as an LMI via the KYP lemma.  
The obtained NTF is an FIR filter, 
which is favorable from the implementation viewpoint.
Design examples have shown effectiveness of our method.

Future work includes STF optimization
as in~\cite{Ho+06}, or
adaptive quantization
as in~\cite{OstZam09}
combined with the proposed optimal filter.

\section*{Acknowledgments}
This research is supported in part by the JSPS Grant-in-Aid
for Scientific Research (B) No.\ 2136020318360203,
Grant-in-Aid for Exploratory Research No.\ 22656095,
and the MEXT Grant-in-Aid for Young Scientists
(B) No.\ 22760317.

\appendix{}

\subsection{Proof of Proposition~\ref{prop:R}}
\label{ap:lemma1}

In this proof, we adopt a standard technique of control theory~\cite{DoyFraTan}.

First assume that $H_1(z)$ and $H_2(z)$ are given by \eqref{eq:param}
for some $P(z) \in {\mathcal{S}}$ and $R(z) \in {\mathcal{S}}'$.
Since $R(z) \in {\mathcal{S}}'$, $R(z)$ is strictly causal and so is
$H_2(z) = R(z)/(1+R(z))$. This implies that the system is well-posed.
For internal stability, we need to show that the four transfer functions
$1/(1-H_2(z))$, $H_1(z)/(1-H_2(z))$, and $H_2(z)/(1-H_2(z))$
are all stable (i.e., their poles are inside the unit circle in the complex plane).
By the equalities in (1), we have
$1/(1-H_2(z))=1+R(z) \in {\mathcal{S}}$,
and hence
$H_1(z)/(1-H_2(z))=P(z)$ and $H_2(z)/(1-H_2(z))=R(z)$
are stable.

Next assume that the feedback system is well-posed and internally stable.
Define
$R := H_2/(1-H_2)$ and $P := H_1/(1-H_2)$.
Since $H_2(z)$ is strictly proper from the well-posedness, 
so is $R(z)$.
Then by the internal stability of the feedback system,
$R=H_2/(1-H_2)$ and $P=H_1/(1-H_1)$ are stable,
that is $R(z)\in{\mathcal{S}}'$ and $P(z)\in{\mathcal{S}}$.
\hfill\IEEEQED
\subsection{Real-valued LMI for Theorem~\ref{thm:gkyp_bp}}
\label{ap:gkyp_bp}

For a Hermitian matrix $F\in\C^{n\times n}$
the inequality $F<0$ 
is equivalent to (\cite{BoyVan})
\[
 \begin{bmatrix}\re F &-\im F\\\im F & \re F\end{bmatrix}<0. 
\]
Hence we obtain the following real-valued LMI for \eqref{eq:gkyp_bp}:
\[
 \begin{bmatrix}M_r(X,Y,\alpha) & -M_i(Y)\\M_i(Y) & M_r(X,Y,\alpha)\end{bmatrix}<0,
\]
where
\[
 \begin{split}
  M_r(X,Y,\alpha) &:= \begin{bmatrix}
		M_{r1}(X,Y) & M_{r2}(X,Y) & C(\alpha)\tr\\
		M_{r2}(X,Y)\tr & M_{r3}(X,\gamma) & 1\\
		C(\alpha) & 1 & -1
	\end{bmatrix},\\
  M_{r1}(X,Y) &:= A\tr XA + (A\tr Y+YA)\cos\omega_0\\&\qquad - X -2Y\cos\Omega,\\
  M_{r2}(X,Y) &:= A\tr XB + YB\cos\omega_0,\\
  M_{r3}(X,\gamma) &:= B\tr X B - \gamma^2,\\
  M_i(Y) &:= \begin{bmatrix}
    M_{i1}(Y)&M_{i2}(Y)&0\\-M_{i2}\tr & 0 & 0\\0&0&0
   \end{bmatrix},\\
  M_{i1}(Y) &:= (A\tr Y - YA)\sin\omega_0,\\
  M_{i2}(Y) &:= -YB\sin\omega_0.
 \end{split}
\]
\hfill\IEEEQED

\subsection{MATLAB codes for optimal NTF}
\label{ap:matlab}

We here introduce MATLAB codes for executing numerical computation
of the design proposed in this paper.
The codes are downloadable from the following web site:

{\footnotesize
\begin{verbatim}
http://www-ics.acs.i.kyoto-u.ac.jp/~nagahara/ds/
\end{verbatim}
}

This site provides a MATLAB function \verb=NTF_MINMAX=, 
which is the main function to design optimal modulators.
Note also that to execute the codes in this section,
Control System Toolbox~\cite{CT}, YALMIP~\cite{Lof04}, and SeDuMi~\cite{Stu01} are needed.
We use Control System Toolbox for defining state-space representation of systems.
YALMIP is a parser for LMI description and SeDuMi
is a solver for convex optimization problem including
LMI's with the self-dual embedding technique.
This function computes the optimal NTF and $R(z)$
minimizing $\gamma>0$ subject to LMI \eqref{eq:gkyp}
for lowpass modulators and \eqref{eq:gkyp_bp} for bandpass modulators.
The $H^\infty$-norm condition of the NTF and assignment of the NTF zeros can be also included
using Lemma~\ref{lem:hinf}.

For example, the optimal lowpass NTF shown in Section~\ref{subsec:example_lp}
is obtained by
\begin{verbatim}
[ntf2,R]=NTF_MINMAX(32,32,1.5^(1/2),0,0);
ntf=ntf2^2;
\end{verbatim}
The optimal bandpass NTF with zeros at $z=\ee^{\pm \jj\pi/2}$ shown in Section~\ref{subsec:example_bp}
is obtained by
\begin{verbatim}
[ntf,R]=NTF_MINMAX(32,16,1.5,1/4,1);
\end{verbatim}

For the optimal multi-band bandpass NTF shown in Section~\ref{subsec:example_bp} is also obtained by
using another function \verb=NTF_MINMAX_MB= as
\begin{verbatim}
ff=[1/8,1/4,3/8];
[ntf,R]=NTF_MINMAX_MB(32,64,1.5,ff,1);
\end{verbatim}

\begin{rem}
When one runs the codes, a message ``{\tt Run into numerical problems}''
may appear.
This means that there was some kind of a numerical problem
encountered in optimization,
and the usefulness of the returned solution should be judged
by the designer.
This may happen occasionally in numerical LMI optimization.
For example, in numerical optimization with an LMI condition $M>0$,
the minimum eigenvalue of $M$ may be slightly negative
due to numerical problems.
In many cases, this does not matter.
To avoid this, one can adopt very small $\varepsilon>0$ and
rewrite $M>0$ as $M>\varepsilon I$.
\end{rem}



\end{document}